\documentclass[prd,twocolumn,nopacs,floatfix,amsmath,nofootinbib,amssymb,floatfix]{revtex4}
\usepackage{graphicx,color,dcolumn,booktabs,bm}
\usepackage{longtable}
\usepackage{pdflscape}
\usepackage{pdfpages}
\usepackage{txfonts}
\usepackage{overpic}
\usepackage{amssymb}
\usepackage{makecell}
\usepackage{indentfirst}
\usepackage{feynmf}   %{feynmp}
\usepackage{slashed}  %for Feynman symbols
\usepackage{cases}
\usepackage{color}
\usepackage{multirow}
\usepackage{threeparttable}
\usepackage{epstopdf}
\usepackage{enumerate}
\usepackage{ulem}
\usepackage[figuresright]{rotating}
\usepackage[colorlinks,
            citecolor=blue,
            anchorcolor=red,
            menucolor=red,
            linkcolor=red,
            filecolor=red,
            runcolor=red,
            urlcolor=blue,
            frenchlinks=true]{hyperref}

\graphicspath{{Figures/}} %

\allowdisplaybreaks

\begin{document}
%%%%%%%%%%%%%%%%%%%%%%%%%%%%%%%%%%%%%%%%%%%%%%%%%%%%%%%%%%%%%%%%%%%
\title{Production of high-orbital kaon excited states in the $K^{-}p$ reaction}
%%%%%%%%%%%%%%%%%%%%%%%%%%%%%%%%%%%%%%%%%%%%%%%%%%%%%%%%%%%%%%%%%%%%
\author{Ting-Yan Li$^{1,2,3,4}$}\email{lity2023@lzu.edu.cn}
\author{Zi-Yue Bai$^{1,2,3,4}$}\email{baiziyue@lzu.edu.cn}
\author{Xiang Liu$^{1,2,3,4}$}\email{xiangliu@lzu.edu.cn}
\affiliation{
$^1$School of Physical Science and Technology, Lanzhou University, Lanzhou 730000, China\\
$^2$Lanzhou Center for Theoretical Physics,
Key Laboratory of Theoretical Physics of Gansu Province,
Key Laboratory of Quantum Theory and Applications of MoE,
Gansu Provincial Research Center for Basic Disciplines of Quantum Physics, Lanzhou University, Lanzhou 730000, China\\
$^3$MoE Frontiers Science Center for Rare Isotopes, Lanzhou University, Lanzhou 730000, China\\
$^4$Research Center for Hadron and CSR Physics, Lanzhou University and Institute of Modern Physics of CAS, Lanzhou 730000, China}

\date{\today}
%%%%%%%%%%%%%%%%%%%%
\begin{abstract}
In this work, a systematic investigation of the production of high-orbital-excitation kaons in $K^{-}p$ reactions is carried out within an effective Lagrangian approach. 
The relevant $t$-channel processes are constructed, and the model is calibrated using a single adjustable parameter determined from existing experimental data. 
With this parameter, the measured production cross sections for the $K_3^*(1780)$, $K_2(1820)$, $K_2(1770)$ and  $K_4^*(2045)$ states are successfully reproduced. 
Employing the same framework, the production cross sections for other high-orbital kaons are predicted. 
The results indicate that these states possess sizable cross sections and exhibit characteristically forward-peaked angular distributions, which is a typical feature of $t$-channel exchange, highlighting their great potential for observation in future experiments.
\end{abstract}
%%%%%%%%%%%%%%%%%%%%%%%%%%%%%%%%%%%%%%%%%%%%%%%%%%%%%%%%%%%%%%%%%%%%

\maketitle

\section{INTRODUCTION} 

Although numerous new hadronic states have been observed in collider experiments over the past two decades \cite{Chen:2022asf,Bai:2026atm,Wang:2025dur,BESIII:2023xac,BaBar:2007ceh,Anisovich:2000ut}, it is important to recognize that meson-beam experiments have played a crucial role in establishing the main spectrum of light hadrons as listed in the Review of Particle Physics (RPP) \cite{ParticleDataGroup:2024cfk}. In recent years, combined with these observed new hadronic states, there have also been numerous phenomenological studies of how to produce them in meson beam experiments \cite{Li:2026pho,Wang:2019uwk,Wang:2017qcw,Wang:2023lia}. At present, the importance of meson beam experiments in the study of hadron spectroscopy should not be neglected; rather, it should be especially emphasized \cite{WANG:2025fmh}.

Generally, a meson beam is chosen as either a pion beam or a kaon beam. Pion-nucleon reactions can produce light mesons without strange quarks, while kaon-nucleon reactions can be applied mainly to produce light mesons with strange quarks. As indicated in the title of this work, excited-kaon production in kaon-nucleon reactions is the focus of our study.

The kaon family, a crucial subset of light hadrons distinguished by the presence of a strange quark, exhibits properties markedly different from those of light non-strange mesons. For decades, meson-beam experiments have served as a primary tool for their study, yielding a wealth of experimental data. Examining the known kaon states, we find that low-lying kaons were discovered many decades ago. However, data for high-lying states are much more limited. A clear trend emerges from the experimental data on the resonance parameters—masses and widths—of these kaons: as the orbital quantum number increases, fewer experimental data are available. Consequently, the spectroscopic picture of the kaon family remains incomplete and requires further refinement.

\begin{figure}[htbp]
 \includegraphics[width=1
    \linewidth]{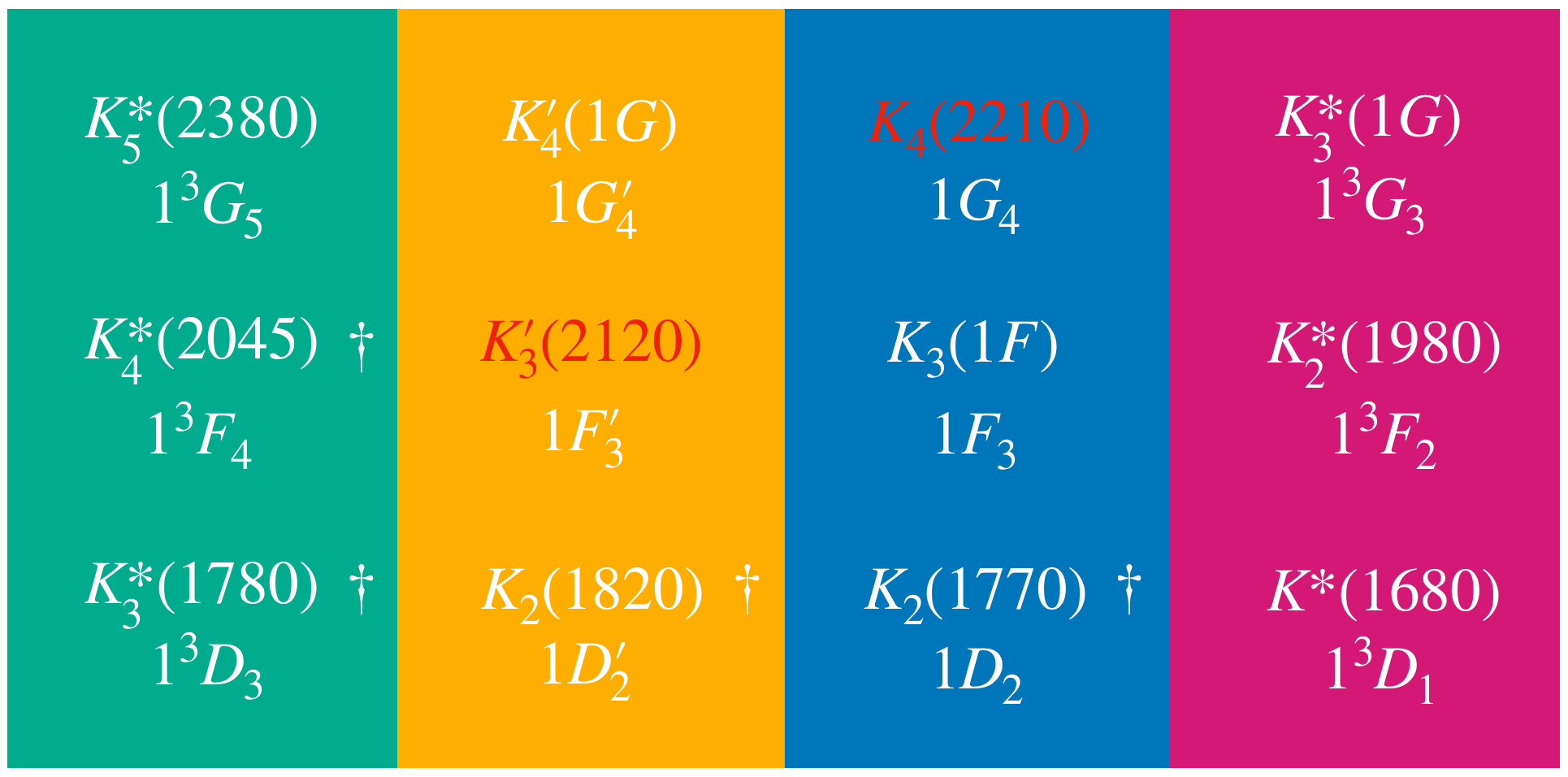}
\caption{A collection of high-orbital kaons. Two newly observed kaons, $K_3^{\prime}(2120)$ and $K_4(2210)$, are marked in red, as reported by the COMPASS Collaboration in 2025 \cite{COMPASS:2025wkw}. Four states marked with asterisks have corresponding experimental cross-section data available for $K^- p$ reactions.}
\label{regge}
\end{figure}

Despite this incomplete picture, the recent discovery of two new kaon resonances, $K_3'(2120)$ and $K_4(2210)$ (highlighted in red in Fig.~\ref{regge}), by the COMPASS Collaboration in 2025 via $Kp$ scattering \cite{COMPASS:2025wkw}, has further enriched the spectrum.
The newly observed $K_3'(2120)$ and $K_4(2210)$ belong to the $1F$ and $1G$ states, respectively. Other well-established high-orbital kaons, such as the $1D$ states $K_3^*(1780)$, $K_2(1820)$, $K_2(1770)$ and the $1F$ state $K_4^*(2045)$, have also been observed in meson-beam experiments, providing clear measurements of their production cross sections \cite{Bird:1988qp,Birmingham-CERN-Glasgow-MichiganState-Paris:1984ppi,Toaff:1981yk,Aston:1993qc,Birmingham-CERN-Glasgow-MichiganState-Paris:1982tev}.
On the theoretical front, significant progress has been made in modeling the mass spectrum via potential models and Regge trajectories \cite{Li:2022khh,Pang:2017dlw}, and in calculating decay properties using frameworks such as the Quark Pair Creation (QPC) model and effective Lagrangian methods \cite{Jafarzade:2022uqo,Jafarzade:2021vhh,Wang:2025lin}.

In stark contrast, the production mechanisms of these high-orbital kaons have received little theoretical attention. Their production dynamics thus represent a vital and underexplored area of study. This paper provides a brief overview of kaon production in meson-beam experiments, which provides an excellent platform for such investigations \cite{WANG:2025fmh}. Future facilities, such as those planned at J-PARC and the potential meson-beam installation at HIAF, underscore the continued relevance of this approach. Building on these experimental prospects and on established theoretical spectroscopy, we calculate the production cross sections for the discussed high-orbital kaons shown in Fig.~\ref{regge} in the following sections. Our calculations are performed using a unified parameter set, based on the robust experimental data available for the $K_3^*(1780)$, $K_2(1820)$ and $K_2(1770)$ states.

The paper is organized as follows. Following this Introduction, we present the effective Lagrangian approach in Section~\ref{section2} to investigate the production of the high-orbital kaons in the $K^- p$ reaction. Numerical results and discussion are presented in Section~\ref{section3}.  In Section~\ref{section4}, we summarize the present work.

\section{Framework for studying high-orbital kaon production via the $K^- p$ reaction}\label{section2}

\begin{figure*}[htbp]
 \includegraphics[width=1.0\textwidth]{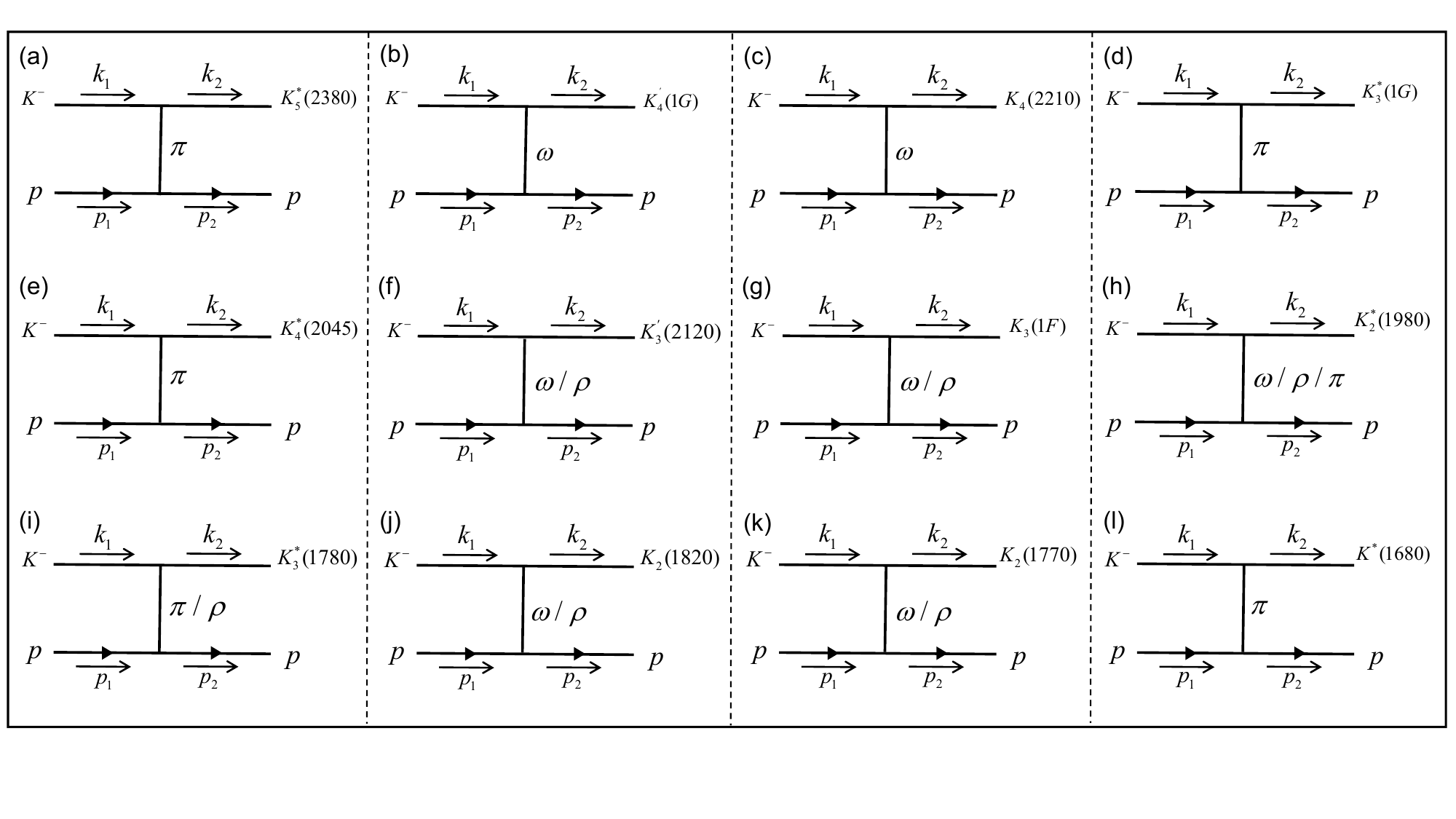}
\caption{Feynman diagrams for the $K^- p \to K_J +N$ reactions.}
\label{feynman}
\end{figure*}

In this section, we present the theoretical foundation required for the subsequent calculations. The Feynman diagrams considered are shown in Fig.~\ref{feynman}. Since the pole mass of the intermediate nucleon is well below the production threshold of the final-state kaon and proton, the contribution from the s-channel is highly suppressed. Moreover, the u-channel contribution is negligible because baryon–antibaryon decay modes of these high-orbital-excitation kaons are either kinematically disfavored or experimentally insignificant. Therefore, we adopt the effective Lagrangian approach to calculate the corresponding $t$-channel amplitude. Since the two-body strong decays of the kaons in Fig.~\ref{regge} have already been calculated by theorists, and their theoretical results are in good agreement with experimental data \cite{Pang:2017dlw,Wang:2025lin}, we consider $t$-channel Feynman diagrams where the exchanged particles originate from kaon-containing decay channels.

\subsection{$1D$-wave states}

For the $1D$-wave states in Fig.~\ref{regge}, the $K_3^*(1780)$ corresponds to the $1^3D_3$ state. The $K_3^*(1780)$ is established in the PDG as a strange meson with $I(J^P) = 1/2\,(3^-)$ strange meson \cite{ParticleDataGroup:2024cfk}. Its resonance parameters are listed with an average mass and width of $1779 \pm 8$ MeV and $161 \pm 17$ MeV, respectively. The mass of the $K_3^*(1780)$ has been measured in numerous production channels over several decades, beginning with its first observations in the 1970s \cite{Baldi:1976ua,Konigs:1978at,Chung:1977ji}. The PDG average mass is derived from a series of experiments conducted in reactions such as $K^{-}p \to K^{-}\pi^{+}n$ and $K^{+}p \to K^{0}\pi^{+}p$ \cite{Aston:1987ir,Baldi:1976ua}. The dominant hadronic decay of the $K_3^*(1780)$ is $K\rho$ with a branching fraction of $(31 \pm 9)\%$, followed by $K^*(892)\pi$ ($(20 \pm 5)\%$) and $K\pi$ ($(18.8 \pm 1.0)\%$) \cite{Aston:1986jb}. The branching ratio for $\pi K_{2}^{*}(1430)$ is experimentally limited to $<16\%$ (CL=95\%) \cite{Aston:1986jb}. Based on the theoretical predictions compiled in Ref. \cite{Wang:2025lin}, the hadronic decay properties of the $K_3^*(1780)$ are further detailed. The dominant channel is $\rho K^*$, which is followed by the $\omega K^*$ and $\pi K$ channels. 

The states $K_2(1770)$ and $K_2(1820)$ correspond to mixtures\footnote{Following Ref.~\cite{Pang:2017dlw}, for $nL=1D,1F,1G$, the mixing between the spin-singlet and spin-triplet kaon states is defined as $|K(nL)\rangle=\cos\theta_{nL}|n^1L_L\rangle+\sin\theta_{nL}|n^3L_L\rangle$ and $|K^{\prime}(nL)\rangle=-\sin\theta_{nL}|n^1L_L\rangle+\cos\theta_{nL}|n^3L_L\rangle$. For the $1D$ wave, this convention gives $|K(1D)\rangle=|K_2(1770)\rangle$, $|K^{\prime}(1D)\rangle=|K_2(1820)\rangle$. The same convention is adopted for the $1F$ and $1G$ mixed pairs.} of the $1^1D_2$ and $1^3D_2$ states.
 These two states share the same quantum numbers $I(J^P) = 1/2(2^-)$. The $K_2(1770)$ has a measured mass and width of $1773 \pm 8$ MeV and $186 \pm 14$ MeV, respectively. The $K_2(1820)$ is heavier and broader, with a mass of $1819 \pm 12$ MeV and a width of $264 \pm 34$ MeV \cite{ParticleDataGroup:2024cfk}.
Both states have been observed in complementary production processes. The $K_2(1770)$ was first established in $K^{-}p$ scattering experiments, such as $K^{-}p \to K^{-}\omega p$ and $K^{-}p \to K^{-} 2\pi p$, conducted at facilities like LASS \cite{Aston:1993qc,ACCMOR:1981yww}. Its existence has been confirmed in modern analyses, notably in the decay $B^{+} \to J/\psi\phi K^{+}$ by the LHCb Collaboration \cite{LHCb:2016axx}. Similarly, the $K_2(1820)$ was also discovered in the LASS experiment through the $K^{-}\omega$ channel \cite{Aston:1993qc} and has been observed in the same $B^{+} \to J/\psi\phi K^{+}$ decay by LHCb \cite{LHCb:2016axx}. The $K_2(1770)$ primarily decays to $K\pi\pi$, $K^{*}(892)\pi$, and $K_2^{*}(1430)\pi$. The $K_2(1820)$, while sharing these dominant modes, also shows a significant branching fraction to $K\phi$, a channel not prominent for the $K_2(1770)$.
In Ref.~\cite{Pang:2017dlw}\footnote{In Ref.~\cite{Pang:2017dlw}, the decay channel written as $K\pi$ in the discussion of the $K_2(1770)$ decay should read $K_1\pi$; this appears to be a typographical error.}, the strong decay widths of these two states as a function of the mixing angle are provided, and it is predicted that their mixing angle tends to be negative.
The largest $\pi$-containing decay channels for both the $K_2(1770)$ and $K_2(1820)$ mesons are $K \rho$ and $K\omega$. Our calculated cross sections for these two states at $\theta_{1D}=-30^\circ$
are in good agreement with the experimental data, as will be shown and discussed in a later section.

The $K^*(1680)$ corresponds to the $1^3D_1$ state. The $K^*(1680)$ has an average mass and width of $1718 \pm 18$ MeV and $320 \pm 110$ MeV. It was first observed in the 1970s and 1980s in formation experiments such as $K^{-}p \to K^{-}\pi^{+}n$ and $K^{-}p \to \overline{K}^{0}\pi^{+}\pi^{-}n$ conducted with the LASS spectrometer \cite{Aston:1987ir,Aston:1986jb}. More recently, it was also observed in the decay $B^{+} \to J/\psi\phi K^{+}$ by the LHCb Collaboration with a significance of 8.5$\sigma$ \cite{LHCb:2016axx}. Its dominant decays are to $K\pi$ ($\sim$39\%), $K\rho$ ($\sim$31\%), and $K^*(892)\pi$ ($\sim$30\%). The strong decay widths of this state are predicted in Ref. \cite{Pang:2017dlw}.

To study the four production reactions involving the $1D$-wave strange-meson
states, we first construct the effective interaction Lagrangians associated
with the upper vertices in Fig.~\ref{feynman}. These vertices describe the
couplings of the produced strange mesons to the incident kaon and the exchanged
light meson. The interaction structures are chosen to satisfy Lorentz
covariance and to be consistent with the spin-parity quantum numbers of the
participating mesons.

Based on the decay properties summarized in
Refs.~\cite{Pang:2017dlw,Wang:2025lin}, the exchanged mesons retained for the
production of the individual states are selected as follows. For the
$K_3^*(1780)$, both $\pi$ and $\rho$ exchanges are included, whereas the
$\eta$-exchange contribution is neglected because of its very small production
cross section. For the $K_2(1820)$ and $K_2(1770)$, whose important decay modes
involve vector mesons, only $\rho$ and $\omega$ exchanges are considered. The
$\eta$-exchange contribution to the $K_2(1820)$ production is again found to be
negligible. For the $K^*(1680)$, the large branching fractions into $K\pi$ and
$K\rho$ motivate the consideration of both pseudoscalar- and vector-meson
exchange mechanisms. However, since the $\rho$-exchange contribution is
numerically very small, only the $\pi$-exchange contribution is retained in
the final results.

For the spin-three state $K_3^*(1780)$, represented by the rank-three tensor
field $K_3^{*,\mu\nu\rho}$, the $K\pi$ interaction contains three derivatives,
as required by the angular-momentum structure of the vertex. The corresponding
effective Lagrangian is written as
\begin{eqnarray}
\mathcal{L}_{K_3^*(1780)K\pi}
&=&
g_{K_3^*(1780)K\pi}\,
K_3^{*+,\mu\nu\rho}
\left[
K^-\,(\partial_\mu\partial_\nu\partial_\rho\pi^0)
\right.
\nonumber\\
&&\left.
-\pi^0\,(\partial_\mu\partial_\nu\partial_\rho K^-)
\right],
\label{eq:k3starpik}
\end{eqnarray}
where $g_{K_3^*(1780)K\pi}$ denotes the effective coupling constant of the
$K_3^*(1780)K\pi$ vertex.

The interaction of the $K_3^*(1780)$ with a kaon and a vector $\rho$ meson is
constructed with the Levi-Civita tensor. This form accounts for the parity
properties of the $K_3^*(1780)K\rho$ vertex and is given by
\begin{eqnarray}
\mathcal{L}_{K_3^*(1780)K\rho}
&=&
g_{K_3^*(1780)K\rho}\,
\varepsilon^{\mu\nu\rho\sigma}
K_3^{*+,\mu\alpha\beta}
(\partial_\nu\rho^0_{\rho})
\nonumber\\
&&
\times(\partial^\alpha\partial^\beta\partial_\sigma K^-).
\label{eq:k3starkrho}
\end{eqnarray}

The physical states $K_2(1820)$ and $K_2(1770)$ have the same quantum numbers
$I(J^P)=1/2(2^-)$. Since they are
mixtures of the $1^3D_2$ and $1^1D_2$ configurations, the same Lorentz
structures are adopted for their couplings to $K\omega$ and $K\rho$, while
their effective coupling constants are treated independently.

For the $K_2(1820)$ state, the relevant interaction Lagrangians are
\begin{equation}
\mathcal{L}_{K_2(1820)\omega K}
=
g_{K_2(1820)\omega K}\,
K_2^{+,\alpha\beta}\,
\omega_\alpha\partial_\beta K^-,
\label{eq:k21820omegak}
\end{equation}
and
\begin{equation}
\mathcal{L}_{K_2(1820)\rho K}
=
g_{K_2(1820)\rho K}\,
K_2^{+,\alpha\beta}\,
\rho^0_\alpha\partial_\beta K^-.
\label{eq:k21820rhok}
\end{equation}

Analogously, the interactions of the $K_2(1770)$ state with the $\omega K$ and
$\rho K$ channels are described by
\begin{equation}
\mathcal{L}_{K_2(1770)\omega K}
=
g_{K_2(1770)\omega K}\,
K_2^{+,\alpha\beta}\,
\omega_\alpha\partial_\beta K^-,
\label{eq:k21770omegak}
\end{equation}
and
\begin{equation}
\mathcal{L}_{K_2(1770)\rho K}
=
g_{K_2(1770)\rho K}\,
K_2^{+,\alpha\beta}\,
\rho^0_\alpha\partial_\beta K^-.
\label{eq:k21770rhok}
\end{equation}

For the vector state $K^*(1680)$, the $K^*(1680)K\pi$ interaction is described
by the conventional derivative coupling among one vector and two pseudoscalar
mesons. The corresponding Lagrangian reads
\begin{equation}
\mathcal{L}_{K^{*}(1680)K\pi}
=
g_{K^{*}(1680)K\pi}\,
K^{*0,\alpha}
\left(
K^-\partial_\alpha\pi^+
-\pi^+\partial_\alpha K^-
\right).
\label{eq:kstar1680kpi}
\end{equation}

We next specify the interaction Lagrangians at the lower vertices of
Fig.~\ref{feynman}, which describe the couplings of the exchanged light mesons
to the nucleons. For the vector-meson exchanges, both vector and tensor
couplings are included. The corresponding phenomenological Lagrangians are
taken from Refs.~\cite{Rijken:2010zzb,Nagels:2015lfa} and are written as
\begin{equation}
\mathcal{L}_{\omega NN}
=
g_{\omega NN}\overline{N}\gamma_\mu\omega^\mu N
+
\frac{f_{\omega NN}}{4m_p}
\overline{N}\sigma_{\mu\nu}
\left(
\partial^\mu\tilde{\omega}^{\nu}
-\partial^\nu\tilde{\omega}^{\mu}
\right)N,
\label{eq:omegaNN}
\end{equation}
and
\begin{equation}
\mathcal{L}_{\rho NN}
=
g_{\rho NN}\overline{N}\gamma_\mu\rho^\mu N
+
\frac{f_{\rho NN}}{4m_p}
\overline{N}\sigma_{\mu\nu}
\left(
\partial^\mu\rho^\nu
-\partial^\nu\rho^\mu
\right)N.
\label{eq:rhoNN}
\end{equation}

For the pseudoscalar-pion exchange, we employ the standard pseudovector
$\pi NN$ interaction,
\begin{equation}
\mathcal{L}_{\pi NN}
=
\frac{f_{\pi NN}}{m_\pi}\,
\overline{N}\gamma_\mu\gamma^5
\partial^\mu\tilde{\pi}\,N.
\label{eq:NNpi}
\end{equation}
Here, $N$ denotes the nucleon field, while $g_{VNN}$ and $f_{VNN}$ represent,
respectively, the vector coupling constants for the vector meson
$V=\rho,\omega$. The quantities $m_p$ and $m_\pi$ denote the proton and pion
masses, respectively.

Combining the upper three-meson vertices with the meson propagators and the
lower meson--nucleon vertices, we obtain the tree-level $t$-channel amplitudes
for the four production reactions. We define the four-momentum transfer as
$q=p_1-p_2=k_2-k_1$ and the Mandelstam variable as $t=q^2$. The momenta $k_1$
and $p_1$ denote the momenta of the incident kaon and nucleon, respectively,
whereas $k_2$ and $p_2$ denote those of the produced strange meson and the
outgoing nucleon. The phenomenological function $F_t(q)$ accounts for the
finite-size effects at the interaction vertices.

For the $K_3^*(1780)$ production through pion exchange, the amplitude is
\begin{equation}
\begin{split}
i\mathcal{M}_{K_3^*(1780)\pi K}
={}&
g_{K_3^*(1780)\pi K}
\frac{f_{\pi NN}}{m_\pi}
\bar{u}(p_2)\gamma_\alpha\gamma_5u(p_1)\,
q^\alpha
\frac{i}{t-m_\pi^2}
\\
&\times
\epsilon^{*,\mu\nu\rho}(k_2)
\left(
q_\mu q_\nu q_\rho
+k_{1\mu}k_{1\nu}k_{1\rho}
\right)
F_t(q).
\end{split}
\label{eq:1780pi}
\end{equation}

The corresponding $\rho$-exchange amplitude is
\begin{equation}
\begin{split}
i\mathcal{M}_{K_3^*(1780)\rho K}
={}&
-g_{K_3^*(1780)\rho K}
\varepsilon^{\mu\nu\rho\sigma}
\epsilon^*_{\mu\alpha\beta}(k_2)
\frac{-i\left(
g^\mu_{\ \alpha}
+q^\mu q_\alpha/m_\rho^2
\right)}{t-m_\rho^2}
\\
&\times
\bar{u}(p_2)
\left[
g_{\rho NN}\gamma_\xi
-\frac{i f_{\rho NN}}{2m_p}
\sigma_{\delta\xi}q^\delta
\right]
u(p_1)
\\
&\times
q_\nu
k_1^\alpha k_1^\beta k_{1\sigma}.
\end{split}
\label{eq:1780rho}
\end{equation}

For the $K_2(1820)$ production, the $\omega$-exchange contribution is given by
\begin{equation}
\begin{split}
i\mathcal{M}_{K_2(1820)\omega K}
={}&
i\,g_{K_2(1820)\omega K}
\bar{u}(p_2)
\left(
g_{\omega NN}\gamma_\mu
-\frac{i f_{\omega NN}}{2m_p}
\sigma_{\nu\mu}q^\nu
\right)
u(p_1)\,
\\
&\times
\epsilon^{*,\alpha\beta}(k_2)
k_{1\beta}
\frac{-i\left(
g^\mu_{\ \alpha}
+q^\mu q_\alpha/m_\omega^2
\right)}
{t-m_\omega^2}
F_t(q).
\end{split}
\label{eq:1820omega}
\end{equation}

Likewise, the $\rho$-exchange amplitude for the $K_2(1820)$ production reads
\begin{equation}
\begin{split}
i\mathcal{M}_{K_2(1820)\rho K}
={}&
i\,g_{K_2(1820)\rho K}
\bar{u}(p_2)
\left(
g_{\rho NN}\gamma_\mu
-\frac{i f_{\rho NN}}{2m_p}
\sigma_{\nu\mu}q^\nu
\right)
u(p_1)\,
\\
&\times
\epsilon^{*,\alpha\beta}(k_2)
k_{1\beta}
\frac{-i\left(
g^\mu_{\ \alpha}
+q^\mu q_\alpha/m_\rho^2
\right)}
{t-m_\rho^2}
F_t(q).
\end{split}
\label{eq:1820rho}
\end{equation}

The $K_2(1770)$ amplitudes have the same Lorentz structures as those of the
$K_2(1820)$, but involve the coupling constants associated with the
$K_2(1770)$ state. The $\omega$-exchange amplitude is
\begin{equation}
\begin{split}
i\mathcal{M}_{K_2(1770)\omega K}
={}&
i\,g_{K_2(1770)\omega K}
\bar{u}(p_2)
\left(
g_{\omega NN}\gamma_\mu
-\frac{i f_{\omega NN}}{2m_p}
\sigma_{\nu\mu}q^\nu
\right)
u(p_1)\,
\\
&\times
\epsilon^{*,\alpha\beta}(k_2)
k_{1\beta}
\frac{-i\left(
g^\mu_{\ \alpha}
+q^\mu q_\alpha/m_\omega^2
\right)}
{t-m_\omega^2}
F_t(q),
\end{split}
\label{eq:1770omega}
\end{equation}
whereas the corresponding $\rho$-exchange amplitude is
\begin{equation}
\begin{split}
i\mathcal{M}_{K_2(1770)\rho K}
={}&
i\,g_{K_2(1770)\rho K}
\bar{u}(p_2)
\left(
g_{\rho NN}\gamma_\mu
-\frac{i f_{\rho NN}}{2m_p}
\sigma_{\nu\mu}q^\nu
\right)
u(p_1)\,
\\
&\times
\epsilon^{*,\alpha\beta}(k_2)
k_{1\beta}
\frac{-i\left(
g^\mu_{\ \alpha}
+q^\mu q_\alpha/m_\rho^2
\right)}
{t-m_\rho^2}
F_t(q).
\end{split}
\label{eq:1770rho}
\end{equation}

Finally, the $K^*(1680)$ production is dominated by the pion-exchange
mechanism. Its tree-level amplitude is expressed as
\begin{equation}
\begin{split}
i\mathcal{M}_{K^*(1680)\pi K}
={}&
-g_{K^*(1680)\pi K}
\frac{f_{\pi NN}}{m_\pi}
\bar{u}(p_2)\gamma_\mu\gamma_5u(p_1)\,
q^\mu
\frac{i}{t-m_\pi^2}
\\
&\times
\epsilon^{*,\alpha}(k_2)
\left(
q_\alpha+k_{1\alpha}
\right)
F_t(q).
\end{split}
\label{eq:1680pi}
\end{equation}

In the above expressions, $\epsilon^{\mu\nu\rho}(k_2)$,
$\epsilon^{\mu\nu}(k_2)$, and $\epsilon^\mu(k_2)$ denote the polarization
tensor or polarization vector of the produced spin-three, spin-two, and
spin-one strange mesons, respectively. These amplitudes constitute the basic
ingredients used below to calculate the total and differential production
cross sections.

\subsection{$1F$-wave states}
 The $K_4^*(2045)$ is assigned as the $1^3F_4$ state with $I(J^P)=1/2\,(4^+)$ \cite{ParticleDataGroup:2024cfk}. 
The $K_4^*(2045)$ is given an average mass and width of $2048^{+8}_{-9}$ MeV and $199^{+27}_{-19}$ MeV, respectively \cite{ParticleDataGroup:2024cfk}. It has been observed in several production channels. The most precise recent measurement comes from the BESIII experiment via $J/\psi \to K^+K^-\pi^0$, reporting a mass of $2090 \pm 9^{+11}_{-29}$ MeV \cite{BESIII:2019apb}. Earlier studies in the 1980s, such as the LASS experiment using the reaction $K^- p \to K^-\pi^+ n$ at 11 GeV/c, were crucial for its initial identification and the determination of its quantum numbers \cite{Aston:1981th, Aston:1986rm}. The $K_4^*(2045)$ predominantly decays into hadronic final states. The $K\pi$ channel has a branching fraction of $(9.9 \pm 1.2)\%$, while the $K^*(892)\pi\pi$ and $\rho K\pi$ modes have comparable strengths of $(9 \pm 5)\%$ and $(5.7 \pm 3.2)\%$, respectively \cite{Birmingham-CERN-Glasgow-MichiganState-Paris:1982tev,Aston:1987ir}. The theoretical decay pattern of the $K_4^*(2045)$ is calculated in Ref. \cite{Wang:2025lin}. The dominant mode for the $K_4^*(2045)$ is $K^*\rho$. Other significant channels include $\omega K^*$, $\pi K^*$, $K\rho$ , and $\pi K$.

The $K_3^{\prime}(2120)$ and $K_3(1F)$ are mixtures of the $1^3F_3$ and $1^1F_3$ states.
The COMPASS Collaboration reported the observation of two new strange mesons in 2025 via the scattering reaction $K^{-}p \to K^{-} \pi^+\pi^- p$ \cite{COMPASS:2025wkw}. The observed states are the $K_3^{\prime}(2120)$ and $K_4(2210)$, with masses and widths of  $2119 \pm 13^{+45}_{-12}$ MeV,  $270 \pm30^{+40}_{-30}$ MeV and $2210 \pm40^{+80}_{-30}$ MeV, $250\pm 70^{+50}_{-70}$ MeV, respectively. The $K_3(1F)$ is the partner state of the $K_3^{\prime}(2120)$, and has not been observed experimentally. The mass and strong decays of the  $K_3^{\prime}(2120)$ and $K_3(1F)$ are calculated in Ref. \cite{Wang:2025lin}. The mass and total decay width for the $K_3(1F)$ are calculated to be 2066~MeV and 248~MeV, respectively  \cite{Wang:2025lin}. While for the $K_3'(2120)$, its  total decay width is 277~MeV. The latter is in good agreement with the experimental width of $270 \pm 30_{-30}^{+40}$~MeV reported by the COMPASS Collaboration \cite{COMPASS:2025wkw}. The decay of the $K_3(1F)$ is dominated by the $K a_2$ channel. Other significant channels include $\rho K^*$ and $\rho K_1$. The $K_3'(2120)$ decays mainly to $\pi K_3^*(1780)$ and $\pi K_2^*$. Channels like $\rho K^*$ and $K\rho$ are also present but with smaller branching fractions.

The $K_2^*(1980)$ state corresponds to the $1^3F_2$ state. The PDG lists its average mass and width as $1990^{+60}_{-50}$~MeV and $348^{+50}_{-30}$~MeV, respectively \cite{ParticleDataGroup:2024cfk}. This state has been observed in multiple production channels. The most recent and precise measurement comes from the BESIII experiment via the decay $\psi(2S) \to K^{+}K^{-}\eta$, reporting a mass of $2046^{+17+67}_{-16-15}$~MeV \cite{BESIII:2019dme}. An earlier measurement by the LASS Collaboration, using the reaction $K^{-}p \to K^{0}\pi^{+}\pi^{-}n$ at 11~GeV/c, found a mass of $1973 \pm 8 \pm 25$~MeV \cite{Aston:1986jb}. A signal consistent with the $K_2^*(1980)$ was also reported by LHCb in an amplitude analysis of $B^{+} \to J/\psi\phi K^{+}$ decays \cite{LHCb:2016axx}, although a later analysis with higher statistics could not confirm it with significant certainty. The primary decay modes of the $K_2^*(1980)$ include $K_1(1270)\pi$, $K_2(1270)\pi$, $Ka_1$, $Kb_1$, and $Kh_1$, as listed in Ref. \cite{Pang:2017dlw}.

To study the four production reactions involving the $1F$-wave
strange-meson states, we construct the effective interaction
Lagrangians associated with the upper vertices in
Fig.~\ref{feynman}. These vertices describe the couplings of the
produced strange mesons to the incident kaon and the exchanged light
meson. As in the discussion of the $1D$-wave states, the interaction
structures are chosen to satisfy Lorentz covariance and to be
consistent with the spin-parity quantum numbers of the participating
mesons. The meson--nucleon interactions at the lower vertices are the
same as those introduced above and are taken from
Refs.~\cite{Rijken:2010zzb,Nagels:2015lfa}.

Based on the decay properties given in
Refs.~\cite{Pang:2017dlw,Wang:2025lin}, we examine the possible
$t$-channel exchange mechanisms for the individual $1F$-wave states
and retain the numerically relevant contributions. For the
$K_4^*(2045)$ production, the pion-exchange mechanism gives the
dominant contribution, while the $\rho$-, $\omega$-, and
$b_1$-exchange contributions are found to be very small. Therefore,
only the $\pi$ exchange is retained. For the two spin-3 states,
the $K_3^\prime(2120)$ and $K_3(1F)$, the $\rho$- and $\omega$-exchange
mechanisms are included. The $a_1$-exchange contribution to the
$K_3^\prime(2120)$ production and the $b_1$-exchange contribution to
the $K_3(1F)$ production are numerically negligible. For the
$K_2^*(1980)$ production, although the axial-vector channels such as
$K a_1$, $K b_1$, and $K h_1$ can have sizable decay fractions, the
corresponding exchange contributions to the production cross section
are strongly suppressed. By contrast, the $K\pi$ channel leads to the
dominant production contribution. For completeness, the $K_2^*(1980)K\omega$ and
$K_2^*(1980)K\rho$ interactions are also listed below to document the
vector-meson exchange mechanisms examined in the calculation.

For the spin-four state $K_4^*(2045)$, represented by the rank-four
tensor field $K_4^{*,\mu\nu\rho\sigma}$, the $K\pi$ interaction
contains four derivatives, as required by the angular-momentum
structure of the vertex. The corresponding effective Lagrangian is
written as

\begin{equation}
\begin{aligned}
\mathcal{L}_{K_4^{*}(2045)\pi K} ={}&
g_{K_4^{*}(2045)\pi K}
K_4^{*+,\mu\nu\rho\sigma}
\Big[
K^{-}
\big(
\partial_\mu
\partial_\nu
\partial_\rho
\partial_\sigma
\pi^{0}
\big)
\\
&\qquad\qquad
+
\pi^{0}
\big(
\partial_\mu
\partial_\nu
\partial_\rho
\partial_\sigma
K^{-}
\big)
\Big].
\end{aligned}
\label{eq:k4starpik}
\end{equation}

The physical states $K_3^\prime(2120)$ and $K_3(1F)$ are mixtures of
the $1^3F_3$ and $1^1F_3$ configurations. Since they have the same
spin-parity quantum numbers, the same Lorentz structures are adopted
for their couplings to the $K\omega$ and $K\rho$ channels, while the
corresponding effective coupling constants are treated independently.
For the $K_3^\prime(2120)$ state, the relevant interaction
Lagrangians are
\begin{equation}
\begin{aligned}
\mathcal{L}_{K_3^{\prime}(2120)\omega K} ={}&
g_{K_3^{\prime}(2120)\omega K}
K_3^{\prime+,\alpha\beta\gamma}
\omega_\alpha
\partial_\beta
\partial_\gamma
K^- ,
\end{aligned}
\label{eq:k3primeomegak}
\end{equation}
and
\begin{equation}
\begin{aligned}
\mathcal{L}_{K_3^{\prime}(2120)\rho K} ={}&
g_{K_3^{\prime}(2120)\rho K}
K_3^{\prime+,\alpha\beta\gamma}
\rho_\alpha^0
\partial_\beta
\partial_\gamma
K^- .
\end{aligned}
\label{eq:k3primerhok}
\end{equation}

Analogously, the interactions of the $K_3(1F)$ state with the
$\omega K$ and $\rho K$ channels are described by
\begin{equation}
\begin{aligned}
\mathcal{L}_{K_3(1F)\omega K} ={}&
g_{K_3(1F)\omega K}
K_3^{+,\alpha\beta\gamma}
\omega_\alpha
\partial_\beta
\partial_\gamma
K^- ,
\end{aligned}
\label{eq:k3omegak}
\end{equation}
and
\begin{equation}
\begin{aligned}
\mathcal{L}_{K_3(1F)\rho K} ={}&
g_{K_3(1F)\rho K}
K_3^{+,\alpha\beta\gamma}
\rho_\alpha^0
\partial_\beta
\partial_\gamma
K^- .
\end{aligned}
\label{eq:k3rhok}
\end{equation}

For the spin-two state $K_2^*(1980)$, the interactions with a kaon
and a vector meson are constructed using the Levi-Civita tensor. The
corresponding effective Lagrangians are
\begin{equation}
\begin{aligned}
\mathcal{L}_{K_2^*(1980)\omega K} ={}&
g_{K_2^*(1980)\omega K}
\varepsilon^{\mu\nu\lambda\sigma}
K_{2,\mu\alpha}^{*+}
\big(
\partial_\nu \omega_\lambda
\big)
\partial_\sigma
\partial^\alpha
K^- ,
\end{aligned}
\label{eq:k2star1980omegak}
\end{equation}
and
\begin{equation}
\begin{aligned}
\mathcal{L}_{K_2^*(1980)\rho K} ={}&
g_{K_2^*(1980)\rho K}
\varepsilon^{\mu\nu\lambda\sigma}
K_{2,\mu\alpha}^{*+}
\big(
\partial_\nu \rho_\lambda^0
\big)
\partial_\sigma
\partial^\alpha
K^- .
\end{aligned}
\label{eq:k2star1980rhok}
\end{equation}

The $K_2^*(1980)K\pi$ interaction contains two derivatives and is
written as
\begin{equation}
\begin{aligned}
\mathcal{L}_{K_2^{*}(1980)\pi K} ={}&
g_{K_2^{*}(1980)\pi K}
K_2^{*+,\alpha\beta}
\Big[
K^-
\big(
\partial_\alpha
\partial_\beta
\pi^0
\big)
\\
&\qquad
+
\pi^0
\big(
\partial_\alpha
\partial_\beta
K^-
\big)
\Big].
\end{aligned}
\label{eq:k2star1980pik}
\end{equation}

Combining the upper three-meson vertices with the exchanged-meson
propagators and the lower meson--nucleon vertices, we obtain the
tree-level $t$-channel amplitudes for the production of the
$1F$-wave states. The definitions
$q=p_1-p_2=k_2-k_1$ and $t=q^2$ follow those introduced above, and
the phenomenological form factor $F_t(q)$ is included at each
$t$-channel exchange vertex.

For the $K_4^*(2045)$ production through pion exchange, the
amplitude is
\begin{equation}
\begin{aligned}
i\mathcal{M}_{K_4^*(2045)\pi K} ={}&
i g_{K_4^*(2045)\pi K}
\frac{f_{\pi NN}}{m_\pi}
\epsilon^{*,\mu\nu\rho\sigma}(k_2)
q^\alpha
\bar{u}(p_2)
\gamma_\alpha
\gamma_5
u(p_1)
\\
&\times
\frac{i}{t-m_\pi^2}
\Big(
q_\mu q_\nu q_\rho q_\sigma
+
k_{1\mu}k_{1\nu}k_{1\rho}k_{1\sigma}
\Big)
F_t(q).
\end{aligned}
\label{eq:k4star}
\end{equation}

For the $K_3^\prime(2120)$ production, the $\omega$-exchange
contribution is given by
\begin{equation}
\begin{aligned}
i\mathcal{M}_{K_3^{\prime}(2120)\omega K} ={}&
g_{K_3^{\prime}(2120)\omega K}
\bar{u}(p_2)
\left(
g_{\omega NN}\gamma_\mu
-
\frac{i f_{\omega NN}}{2m_p}
\sigma_{\nu\mu}q^\nu
\right)
u(p_1)
\\
&\times
\frac{
-i\left(
g^{\mu}{}_{\alpha}
+
q^\mu q_\alpha/m_\omega^2
\right)
}{
t-m_\omega^2
}
\epsilon^{*,\alpha\beta\gamma}(k_2)
k_{1\beta}
k_{1\gamma}
F_t(q),
\end{aligned}
\label{eq:k3primeomega}
\end{equation}
while the corresponding $\rho$-exchange amplitude reads
\begin{equation}
\begin{aligned}
i\mathcal{M}_{K_3^{\prime}(2120)\rho K} ={}&
g_{K_3^{\prime}(2120)\rho K}
\bar{u}(p_2)
\left(
g_{\rho NN}\gamma_\mu
-
\frac{i f_{\rho NN}}{2m_p}
\sigma_{\nu\mu}q^\nu
\right)
u(p_1)
\\
&\times
\frac{
-i\left(
g^{\mu}{}_{\alpha}
+
q^\mu q_\alpha/m_\rho^2
\right)
}{
t-m_\rho^2
}
\epsilon^{*,\alpha\beta\gamma}(k_2)
k_{1\beta}
k_{1\gamma}
F_t(q).
\end{aligned}
\label{eq:k3primerho}
\end{equation}

The $K_3(1F)$ amplitudes have the same Lorentz structures as those of
the $K_3^\prime(2120)$ state, but involve the coupling constants
associated with the $K_3(1F)$ state. The $\omega$-exchange amplitude
is
\begin{equation}
\begin{aligned}
i\mathcal{M}_{K_3(1F)\omega K} ={}&
g_{K_3(1F)\omega K}
\bar{u}(p_2)
\left(
g_{\omega NN}\gamma_\mu
-
\frac{i f_{\omega NN}}{2m_p}
\sigma_{\nu\mu}q^\nu
\right)
u(p_1)
\\
&\times
\frac{
-i\left(
g^{\mu}{}_{\alpha}
+
q^\mu q_\alpha/m_\omega^2
\right)
}{
t-m_\omega^2
}
\epsilon^{*,\alpha\beta\gamma}(k_2)
k_{1\beta}
k_{1\gamma}
F_t(q),
\end{aligned}
\label{eq:k31fomega}
\end{equation}
whereas the corresponding $\rho$-exchange contribution is
\begin{equation}
\begin{aligned}
i\mathcal{M}_{K_3(1F)\rho K} ={}&
g_{K_3(1F)\rho K}
\bar{u}(p_2)
\left(
g_{\rho NN}\gamma_\mu
-
\frac{i f_{\rho NN}}{2m_p}
\sigma_{\nu\mu}q^\nu
\right)
u(p_1)
\\
&\times
\frac{
-i\left(
g^{\mu}{}_{\alpha}
+
q^\mu q_\alpha/m_\rho^2
\right)
}{
t-m_\rho^2
}
\epsilon^{*,\alpha\beta\gamma}(k_2)
k_{1\beta}
k_{1\gamma}
F_t(q).
\end{aligned}
\label{eq:k31frho}
\end{equation}

The vector-meson exchange amplitudes examined for
the $K_2^*(1980)$ production are given. The $\omega$-exchange
contribution is
\begin{equation}
\begin{aligned}
i\mathcal{M}_{K_2^*(1980)\omega K} ={}&
-i g_{K_2^*(1980)\omega K}
\varepsilon^{\mu\nu\lambda\sigma}
\epsilon_{\mu\alpha}^{*}(k_2)
q_\nu
k_{1\sigma}
k_1^\alpha
\\&\times
\bar{u}(p_2)
\left(
g_{\omega NN}\gamma_\xi
-
\frac{i f_{\omega NN}}{2m_p}
\sigma_{\delta\xi}q^\delta
\right)
u(p_1)
\\&\times
\frac{
-i\left(
g^{\xi}{}_{\lambda}
+
q^\xi q_\lambda/m_\omega^2
\right)
}{
t-m_\omega^2
}
F_t(q),
\end{aligned}
\label{eq:1980omega}
\end{equation}
and the $\rho$-exchange contribution is
\begin{equation}
\begin{aligned}
i\mathcal{M}_{K_2^*(1980)\rho K} ={}&
-i g_{K_2^*(1980)\rho K}
\varepsilon^{\mu\nu\lambda\sigma}
\epsilon_{\mu\alpha}^{*}(k_2)
q_\nu
k_{1\sigma}
k_1^\alpha
\\& \times
\bar{u}(p_2)
\left(
g_{\rho NN}\gamma_\xi
-
\frac{i f_{\rho NN}}{2m_p}
\sigma_{\delta\xi}q^\delta
\right)
u(p_1)
\\&\times
\frac{
-i\left(
g^{\xi}{}_{\lambda}
+
q^\xi q_\lambda/m_\rho^2
\right)
}{
t-m_\rho^2
}
F_t(q).
\end{aligned}
\label{eq:1980rho}
\end{equation}

The $K_2^*(1980)$ production also retains the dominant pion-exchange
contribution. The corresponding amplitude is
\begin{equation}
\begin{aligned}
i\mathcal{M}_{K_2^*(1980)\pi K} ={}&
-i g_{K_2^{*}(1980)\pi K}
\frac{f_{\pi NN}}{m_\pi}
\epsilon^{*,\alpha\beta}(k_2)
\left(
q_\alpha q_\beta
+
k_{1\alpha}k_{1\beta}
\right)
\\&\times
\frac{i}{t-m_\pi^2}
\bar{u}(p_2)
\gamma_\mu
\gamma_5
u(p_1)
q^\mu
F_t(q).
\end{aligned}
\label{eq:1980pi}
\end{equation}

In the above expressions,
$\epsilon^{\mu\nu\rho\sigma}(k_2)$,
$\epsilon^{\mu\nu\rho}(k_2)$, and
$\epsilon^{\mu\nu}(k_2)$ denote the polarization tensors of the
produced spin-four, spin-three, and spin-two strange mesons,
respectively. These amplitudes provide the ingredients required to
evaluate the total and differential cross sections for the $1F$-wave states.

\subsection{$1G$-wave states}

The $K_5^*(2380)$ is a candidate for the $1^3G_5$ state with
$I(J^P)=1/2\,(5^-)$~\cite{ParticleDataGroup:2024cfk}.
Its experimental status remains to be confirmed. Its reported mass
and width are $2382 \pm 24$ MeV and $180 \pm 50$ MeV,
respectively~\cite{ParticleDataGroup:2024cfk}. To date, this state
has only been observed in the LASS analysis of the reaction
$K^- p \to K^- \pi^+ n$~\cite{Aston:1986rm}. The only
experimentally established two-body decay mode is $K\pi$, with a
branching fraction of $(6.1 \pm 1.2)\%$, while theoretical calculations
predict a rich decay pattern with numerous allowed open channels, owing to its relatively high mass and large spin. The
two largest predicted channels are $K^* a_2(1320)$ and
$\rho K_2^*(1430)$. Other important modes include $\rho K^*$,
$K^* f_2(1270)$, $\omega K_2^*$, and
$K b_1$~\cite{Wang:2025lin}.

The $K_4^{\prime}(1G)$ and $K_4(2210)$ are mixtures of the
$1^3G_4$ and $1^1G_4$ states.  Their masses and strong decay
properties were investigated in Ref.~\cite{Wang:2025lin}. Its decay is
dominated by the $K a_2$ channel, with a branching fraction of
$22.7\%$ and a partial width of $26.1$ MeV. Other notable channels
include $\rho K_1$ and $\rho K^*$. In contrast, the predicted mass
and total width of the as-yet-unobserved $K_4^{\prime}(1G)$ are
$2306$ MeV and $161$ MeV, respectively. Its important decay modes
include $\pi K_3^*(1780)$, $\pi K_4^*(2045)$, $\pi K_2^*$,
$\rho K^*$, and $\rho K$.

The $K_3^*(1G)$ is assigned as the $1^3G_3$ state and has not yet
been observed experimentally. We studied its mass and strong decay properties
 using the modified Godfrey--Isgur (MGI) model and the
QPC model, and the parameters are given in Ref. \cite{Wang:2025lin}. Its predicted mass
and total width are $2305$ MeV and $412.04$ MeV, respectively. The
dominant decay modes are expected to be $K_2(1770)\pi$,
$K_2(1820)\pi$, $K\pi_2$, and $K\eta_2$. The $K_1\rho$,
$K^*\rho$, and $K b_1$ channels also provide sizable contributions.

To study the production of the four $1G$-wave kaon states,
we construct the effective interaction Lagrangians associated with
the upper vertices in Fig.~\ref{feynman}. These vertices describe
the couplings of the produced strange mesons to the incident kaon
and the exchanged light meson. As in the discussions of the
$1D$- and $1F$-wave states, the interaction structures are chosen
to satisfy Lorentz covariance and to be consistent with the
spin-parity quantum numbers of the participating mesons. The
meson--nucleon interactions at the lower vertices are the same as
those introduced above.

Based on the decay properties obtained in
Ref.~\cite{Wang:2025lin}, we examine the possible $t$-channel
exchange mechanisms and retain the numerically relevant
contributions. It should be emphasized that a sizable decay
branching fraction does not necessarily imply a dominant production
contribution, since the exchanged-meson propagator, the
meson--nucleon coupling, and the momentum structure of the vertex
also affect the production amplitude. For the $K_5^*(2380)$
production, the pion-exchange mechanism provides the dominant
contribution, whereas the $\rho$-, $\omega$-, $b_1$-, and
$h_1$-exchange contributions are numerically negligible. Therefore,
only the $\pi$ exchange is retained. For the $K_4^{\prime}(1G)$ and
$K_4(2210)$ states, only the $\omega$-exchange mechanism is
considered, because the corresponding $\rho$-exchange contributions
are strongly suppressed. For the $K_3^*(1G)$ production, although
the axial-vector channels such as $K a_1$, $K b_1$, and $K h_1$
can have sizable decay fractions, their exchange contributions to
the production cross section are small. By contrast, the $K\pi$
channel leads to the dominant production contribution. Therefore,
only the pion-exchange mechanism is retained for the $K_3^*(1G)$
state.

For the spin-five state $K_5^*(2380)$, represented by the rank-five
tensor field $K_5^{*,\alpha\beta\gamma\delta\sigma}$, the $K\pi$
interaction contains five derivatives, as required by the
angular-momentum structure of the vertex. The corresponding
effective Lagrangian is written as
\begin{equation}
\begin{aligned}
\mathcal{L}_{K_5^*(2380)\pi K}
={}&
g_{K_5^*(2380)\pi K}
K_5^{*+,\alpha\beta\gamma\delta\sigma}
\Big[
K^-
\big(
\partial_\alpha \partial_\beta \partial_\gamma
\partial_\delta \partial_\sigma \pi^0
\big)
\\
&\hspace{1.5em}
-
\pi^0
\big(
\partial_\alpha \partial_\beta \partial_\gamma
\partial_\delta \partial_\sigma K^-
\big)
\Big].
\end{aligned}
\label{eq:k5starkpik}
\end{equation}

The physical states $K_4^{\prime}(1G)$ and $K_4(2210)$ have the
same spin-parity quantum numbers. Since they arise from the mixing
of the $1^3G_4$ and $1^1G_4$ configurations, the same Lorentz
structure is adopted for their couplings to the $K\omega$ channel,
while the corresponding effective coupling constants are treated
independently. For the $K_4^{\prime}(1G)$ state, the interaction
Lagrangian is
\begin{equation}
\begin{aligned}
\mathcal{L}_{K_4^{\prime}(1G)\omega K}
={}&
g_{K_4^{\prime}(1G)\omega K}
K_4^{\prime+,\alpha\beta\gamma\delta}
\omega_\alpha
\partial_\beta \partial_\gamma \partial_\delta K^- .
\end{aligned}
\label{eq:k4primepipi1}
\end{equation}

Analogously, the $K_4(2210)K\omega$ interaction is described by
\begin{equation}
\begin{aligned}
\mathcal{L}_{K_4(2210)\omega K}
={}&
g_{K_4(2210)\omega K}
K_4^{+,\alpha\beta\gamma\delta}
\omega_\alpha
\partial_\beta \partial_\gamma \partial_\delta K^- .
\end{aligned}
\label{eq:k42210omegak}
\end{equation}

For the spin-three state $K_3^*(1G)$, represented by the rank-three
tensor field $K_3^{*,\alpha\beta\gamma}$, the $K\pi$ interaction
contains three derivatives. Its effective Lagrangian is written as
\begin{equation}
\begin{aligned}
\mathcal{L}_{K_3^*(1G)\pi K}
={}&
g_{K_3^*(1G)\pi K}
K_3^{*+,\alpha\beta\gamma}
\Big[
K^-
\big(
\partial_\alpha \partial_\beta \partial_\gamma \pi^0
\big)
\\
&\hspace{1.5em}
-
\pi^0
\big(
\partial_\alpha \partial_\beta \partial_\gamma K^-
\big)
\Big].
\end{aligned}
\label{eq:k3starkpik}
\end{equation}

Combining the upper three-meson vertices with the exchanged-meson
propagators and the lower meson--nucleon vertices, we obtain the
tree-level $t$-channel amplitudes for the production of the
$1G$-wave states. We use the same momentum convention as above,
$q=p_1-p_2=k_2-k_1$ and $t=q^2$.

For the $K_5^*(2380)$ production through pion exchange, the
amplitude is
\begin{equation}
\begin{aligned}
i\mathcal{M}_{K_5^*(2380)\pi K}
={}&
-
g_{K_5^*(2380)\pi K}
\frac{f_{\pi NN}}{m_\pi}
\epsilon^{*,\sigma\alpha\beta\gamma\delta}(k_2)
\frac{i}{t-m_\pi^2}
\\& \times
\Big(
q_\sigma q_\alpha q_\beta q_\gamma q_\delta
+k_{1\sigma} k_{1\alpha} k_{1\beta}
k_{1\gamma} k_{1\delta}
\Big)
\\
&\times
q^\mu
\bar{u}(p_2)\gamma_\mu\gamma_5 u(p_1)
F_t(q).
\end{aligned}
\label{eq:2380}
\end{equation}

For the $K_4^{\prime}(1G)$ production, the retained
$\omega$-exchange contribution is
\begin{equation}
\begin{aligned}
i\mathcal{M}_{K_4^{\prime}(1G)\omega K}
={}&
-i g_{K_4^{\prime}(1G)\omega K}
\bar{u}(p_2)
\Bigg(
g_{\omega NN}\gamma_\mu
-
\frac{i f_{\omega NN}}{2m_p}
\sigma_{\nu\mu}q^\nu
\Bigg)
u(p_1)
\\
&\times
\frac{-i}{t-m_\omega^2}
\Bigg(
g^\mu_{\ \alpha}
+
\frac{q^\mu q_\alpha}{m_\omega^2}
\Bigg)
\epsilon^{*,\alpha\beta\gamma\delta}(k_2)
k_{1\beta}k_{1\gamma}k_{1\delta}
F_t(q).
\end{aligned}
\label{eq:k4prime}
\end{equation}

The $K_4(2210)$ amplitude has the same Lorentz structure, but
involves the coupling constant associated with the $K_4(2210)$
state:
\begin{equation}
\begin{aligned}
i\mathcal{M}_{K_4(2210)\omega K}
={}&
-i g_{K_4(2210)\omega K}
\bar{u}(p_2)
\Bigg(
g_{\omega NN}\gamma_\mu
-
\frac{i f_{\omega NN}}{2m_p}
\sigma_{\nu\mu}q^\nu
\Bigg)
u(p_1)
\\
&\times
\frac{-i}{t-m_\omega^2}
\Bigg(
g^\mu_{\ \alpha}
+
\frac{q^\mu q_\alpha}{m_\omega^2}
\Bigg)
\epsilon^{*,\alpha\beta\gamma\delta}(k_2)
k_{1\beta}k_{1\gamma}k_{1\delta}
F_t(q).
\end{aligned}
\label{eq:k42210}
\end{equation}

Finally, the $K_3^*(1G)$ production is dominated by the
pion-exchange mechanism. The corresponding amplitude is
\begin{equation}
\begin{aligned}
i\mathcal{M}_{K_3^*(1G)\pi K}
={}&
g_{K_3^*(1G)\pi K}
\frac{f_{\pi NN}}{m_\pi}
\bar{u}(p_2)\gamma_\alpha\gamma_5 u(p_1)
q^\alpha
\frac{i}{t-m_\pi^2}
\\
&\times
\epsilon^{*,\mu\nu\rho}(k_2)
\Big(
q_\mu q_\nu q_\rho
+
k_{1\mu} k_{1\nu} k_{1\rho}
\Big)
F_t(q).
\end{aligned}
\label{eq:k3star1G}
\end{equation}

In the above expressions,
$\epsilon^{\mu\nu\rho\sigma\lambda}(k_2)$,
$\epsilon^{\mu\nu\rho\sigma}(k_2)$, and
$\epsilon^{\mu\nu\rho}(k_2)$ denote the polarization tensors of
the produced spin-five, spin-four, and spin-three strange mesons,
respectively. The quantities $\bar{u}(p_2)$ and $u(p_1)$ are the
Dirac spinors of the outgoing and incoming nucleons. These
amplitudes provide the ingredients required to evaluate the total
and differential cross sections for the $1G$-wave states. For $t-$channel meson exchange, the form factor  is adopted as
\begin{equation}
\begin{split}
F_{t}(q) = \left( \frac{\Lambda_{t}^{2} - m^{2}}{\Lambda_{t}^{2} - q^{2}} \right)^{2}, \end{split}
\label{Ft}
\end{equation}
where $m$ represents the mass of the exchanged particle. 
By fitting the available experimental cross section data at different energies for the reaction $K^- p \to K_3^*(1780)p$~\cite{Bird:1988qp,Birmingham-CERN-Glasgow-MichiganState-Paris:1984ppi,Toaff:1981yk}, we obtain $\Lambda_t = 1.5 \pm 0.2$ GeV, with $\chi^2/\mathrm{d.o.f.} \leq 0.4$.
Employing the same cutoff value for the $K^- p \to K_2(1820) p$ and $K^- p \to K_2(1770) p$  reactions, we find that the theoretical predictions are in good agreement with the measurements reported in Refs.~\cite{,Aston:1993qc,Estabrooks:1977xe,Birmingham-CERN-Glasgow-MichiganState-Paris:1982tev}. 

The Regge trajectory model successfully describes hadron production at high energies~\cite{Ozaki:2009wp,Storrow:1983ct}. Replacing the Feynman propagator with a Regge propagator effectively accounts for the exchange of high-spin and high-mass states in the $t$-channel, whose contributions become significant at higher energies. We implement Reggeization by substituting the $t$-channel Feynman propagators with Regge propagators~\cite{Guidal:1997hy}:
\begin{equation}
\frac{1}{t-m_{\pi}^2} \rightarrow 
\left(\frac{s}{s_{\text{scale}}}\right)^{\alpha_{\pi}(t)}
\frac{\pi\alpha_{\pi}^{\prime}}{\Gamma[\alpha_{\pi}(t)+1] \sin[\pi\alpha_{\pi}(t)]},
\label{pi_propagator}
\end{equation}

\begin{equation}
\frac{1}{t-m_{R}^2} \rightarrow 
\left(\frac{s}{s_{\text{scale}}}\right)^{\alpha_{R}(t)-1} 
\frac{\pi\alpha_{R}^{\prime}}{\Gamma[\alpha_{R}(t)] \sin[\pi\alpha_{R}(t)]},
\label{rho_propagator}
\end{equation}
the corresponding Regge trajectories for the exchanged mesons are
\begin{equation}
\alpha_R(t) = 1 + \alpha'_R\left(t - m_R^2\right), 
\qquad R = \rho,\;\omega,\;
\label{alphaR}
\end{equation}

\begin{equation}
\alpha_{\pi}(t) = \alpha_\pi^\prime(t-m_{\pi}^2),
\label{alpha_pi}
\end{equation}
with slopes $\alpha'_\rho=0.8\;\mathrm{GeV}^{-2}$, $\alpha'_\omega=0.9\;\mathrm{GeV}^{-2}$, and $\alpha_\pi^\prime = 0.7\,\text{GeV}^{-2}$ ~\cite{Guidal:1997hy}.

Within the preceding theoretical framework, the cross sections for all reactions can be computed. The differential cross section in the center-of-mass (c.m.) frame is given by

\begin{equation}
\frac{d\sigma}{d\cos\theta_{cm}} = \frac{1}{32\pi s} \frac{\left|\vec{k}_{2}^{\,\mathrm{c.m.}}\right|}{\left|\vec{k}_{1}^{\,\mathrm{c.m.}}\right|} \left( \frac{1}{2} \sum_{\lambda} \left|\mathcal{M}\right|^{2} \right),
\label{diff}
\end{equation}
where $s = (k_{1} + p_{1})^{2}$, and $\theta_{cm}$ denotes the angle between the outgoing meson and the incoming $K^{-}$ beam direction in the c.m. frame. The three-momenta $\vec{k}_{1}^{\,\mathrm{c.m.}}$ and $\vec{k}_{2}^{\,\mathrm{c.m.}}$ refer to the initial $K^{-}$ beam and the final meson, respectively. The factor $1/2$ averages over the initial nucleon spin states.

\section{Numerical results and discussions}
\label{section3}
Using the Lagrangians for the three-meson vertices introduced in the preceding section, we evaluate all relevant coupling constants, which are summarized in Table~\ref{cp}. Decay widths used to determine the coupling constants are listed in Table~\ref{width}. With these couplings and the amplitude formulas given earlier, we compute the total and differential production cross sections for the $1D$-, $1F$-, and $1G$-wave states.

\begin{table}[htbp]
\centering
\caption{
Coupling constants used in this work. The listed three-meson coupling
constants are determined by fitting the decay widths from Refs.~\cite{Wang:2025lin,Pang:2017dlw}.
The meson-nucleon-nucleon coupling constants entering the lower vertices
of the diagrams in Fig.~\ref{feynman} are taken from Ref.~\cite{Rijken:2010zzb}. Couplings with mass
dimensions are given with explicit units, while those without units are
dimensionless. \label{cp}
}

\renewcommand\arraystretch{1.8}
\setlength{\tabcolsep}{4pt}
\begin{tabular}{ll}
\toprule[0.8pt]
$g_{K_3^*(1780)\pi K}=3.84\ \mathrm{GeV}^{-2}$& $g_{K_3^*(1780)\rho K}=12.36\ \mathrm{GeV}^{-3}$ \\ 
$g_{K_2(1820)\omega K}=5.85$&  $g_{K_2(1820)\rho K}=5.70$\\
$g_{K_2(1770)\omega K}=5.93$&  $g_{K_2(1770)\rho K}=5.79$\\
$g_{K^*(1680)\pi K}=1.76$& $g_{K_4^*(2045)\pi K}=3.15\ \mathrm{GeV}^{-3}$\\
$g_{K_3^{\prime}(2120) \omega K}=3.02\ \mathrm{GeV}^{-1}$ &
$g_{K_3^{\prime}(2120) \rho K}=2.97\ \mathrm{GeV}^{-1}$\\
 $g_{K_3(1F) \omega K}=2.71\ \mathrm{GeV}^{-1}$ & $g_{K_3(1F) \rho K}=2.63\ \mathrm{GeV}^{-1}$ \\
 $g_{K_2^*(1980) \omega K}=1.09\ \mathrm{GeV}^{-2}$ &$g_{K_2^*(1980) \rho K}=1.06\ \mathrm{GeV}^{-2}$ \\
 $g_{K_2^*(1980) \pi K}=0.24\ \mathrm{GeV}^{-1}$ & $g_{K_5^*(2380) \pi K}=1.51\ \mathrm{GeV}^{-4}$ \\
$g_{K_4^{\prime}(1G) \omega K}=2.13\ \mathrm{GeV}^{-2}$& $g_{K_4(2210) \omega K}=2.09\ \mathrm{GeV}^{-2}$\\ 
 $g_{K_3^*(1G) \pi K}=0.33\ \mathrm{GeV}^{-2}$\\ 
\bottomrule[0.8pt]
\end{tabular}
\end{table}

\begin{table}[htbp]
\centering
\caption{
Decay widths used to determine the three-meson coupling constants in the present work.
Experimental widths (denoted by subscript ${\rm E}$) are taken from the PDG~\cite{ParticleDataGroup:2024cfk},
while theoretical widths (denoted by subscript ${\rm T}$) are adopted from
Refs.~\cite{Wang:2025lin,Pang:2017dlw} when corresponding experimental data are unavailable.
All widths are given in MeV.
}
\label{width}

\renewcommand{\arraystretch}{1.6}
\setlength{\tabcolsep}{5pt}

\begin{tabular}{ll}
\toprule[0.8pt]

$\Gamma_{K_3^*(1780)\to \pi K}=30.3^{\rm E}$ &
$\Gamma_{K_3^*(1780)\to \rho K}=49.9^{\rm E}$ \\

$\Gamma_{K_2(1820)\to \omega K}=46.2^{\rm T}$ &
$\Gamma_{K_2(1820)\to \rho K}=138.1^{\rm T}$ \\

$\Gamma_{K_2(1770)\to \omega K}=41.9^{\rm T}$ &
$\Gamma_{K_2(1770)\to \rho K}=126.0^{\rm T}$ \\

$\Gamma_{K^*(1680)\to \pi K}=79.3^{\rm E}$ &
$\Gamma_{K_4^*(2045)\to \pi K}=19.7^{\rm E}$ \\

$\Gamma_{K_3^{\prime}(2120)\to \omega K}=6.7^{\rm T}$ &
$\Gamma_{K_3^{\prime}(2120)\to \rho K}=20.6^{\rm T}$ \\

$\Gamma_{K_3(1F)\to \omega K}=4.5^{\rm T}$ &
$\Gamma_{K_3(1F)\to \rho K}=13.5^{\rm T}$ \\

$\Gamma_{K_2^*(1980)\to \omega K}=3.6^{\rm T}$ &
$\Gamma_{K_2^*(1980)\to \rho K}=10.6^{\rm T}$ \\

$\Gamma_{K_2^*(1980)\to \pi K}=0.8^{\rm T}$ &
$\Gamma_{K_5^*(2380)\to \pi K}=9.0^{\rm E}$ \\

$\Gamma_{K_4^{\prime}(1G)\to \omega K}=2.3^{\rm T}$ &
$\Gamma_{K_4(2210)\to \omega K}=1.5^{\rm T}$ \\

$\Gamma_{K_3^*(1G)\to \pi K}=1.04^{\rm T}$ &
\\

\bottomrule[0.8pt]
\end{tabular}

\vspace{1mm}
\end{table}

\subsection{Analysis of $1D$-wave kaon production} 

\begin{figure}[htbp]
 \includegraphics[width=0.42\textwidth]{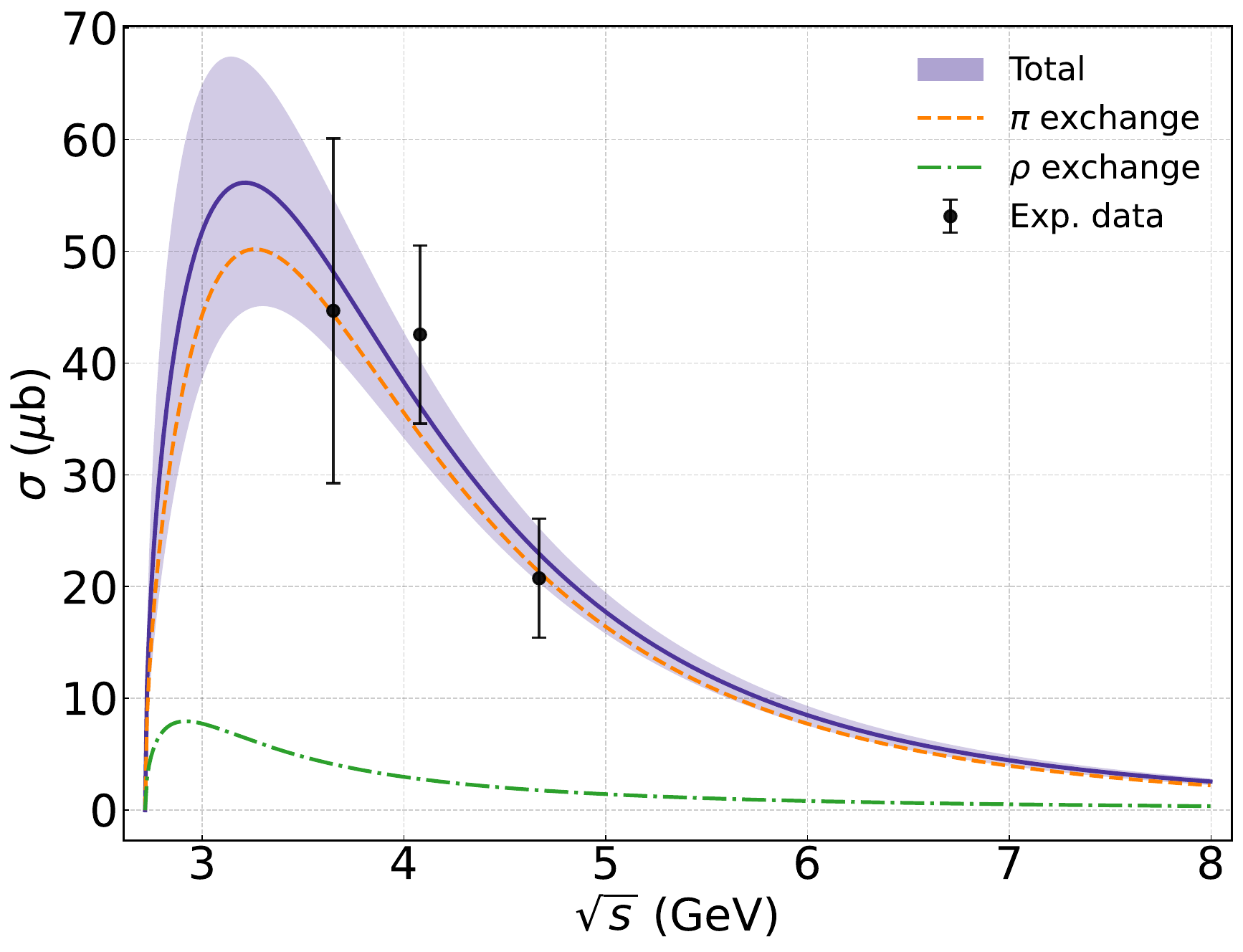}
\caption{The calculated total production cross section for the $K_3^*(1780)$ state is presented, along with comparisons to experimental data. The experimental data are taken from Refs.~\cite{Bird:1988qp,Birmingham-CERN-Glasgow-MichiganState-Paris:1984ppi,Toaff:1981yk}. All error bands and the solid lines within them correspond to results obtained with the cutoff parameter set to $1.5\pm 0.2$ GeV.}
\label{ts_total}
\end{figure}

As shown in Fig.~\ref{ts_total}, we first analyze the total cross section of the
$K^{-}p\to K^{*}_{3}(1780)p$ reaction. In the calculation, both the
$\pi$- and $\rho$-exchange contributions are included, as indicated in
the figure. It can be seen that the $\pi$ exchange gives the dominant
contribution to the total cross section, while the $\rho$ exchange only
provides a relatively small correction. In fitting the experimental total
cross sections, the only adjustable parameter is the cutoff parameter
$\Lambda_t$ in the form factor. The best description of the data is
obtained with $\Lambda_t=1.5~{\rm GeV}$, and the theoretical uncertainty
band is generated by varying this value by $\pm 0.2~{\rm GeV}$. With this
single parameter, the calculated total cross section can reproduce the
available experimental measurements over the considered energy region,
especially around the energy range, where the data are concentrated.

\begin{figure}[htbp]
 \includegraphics[width=0.45\textwidth]{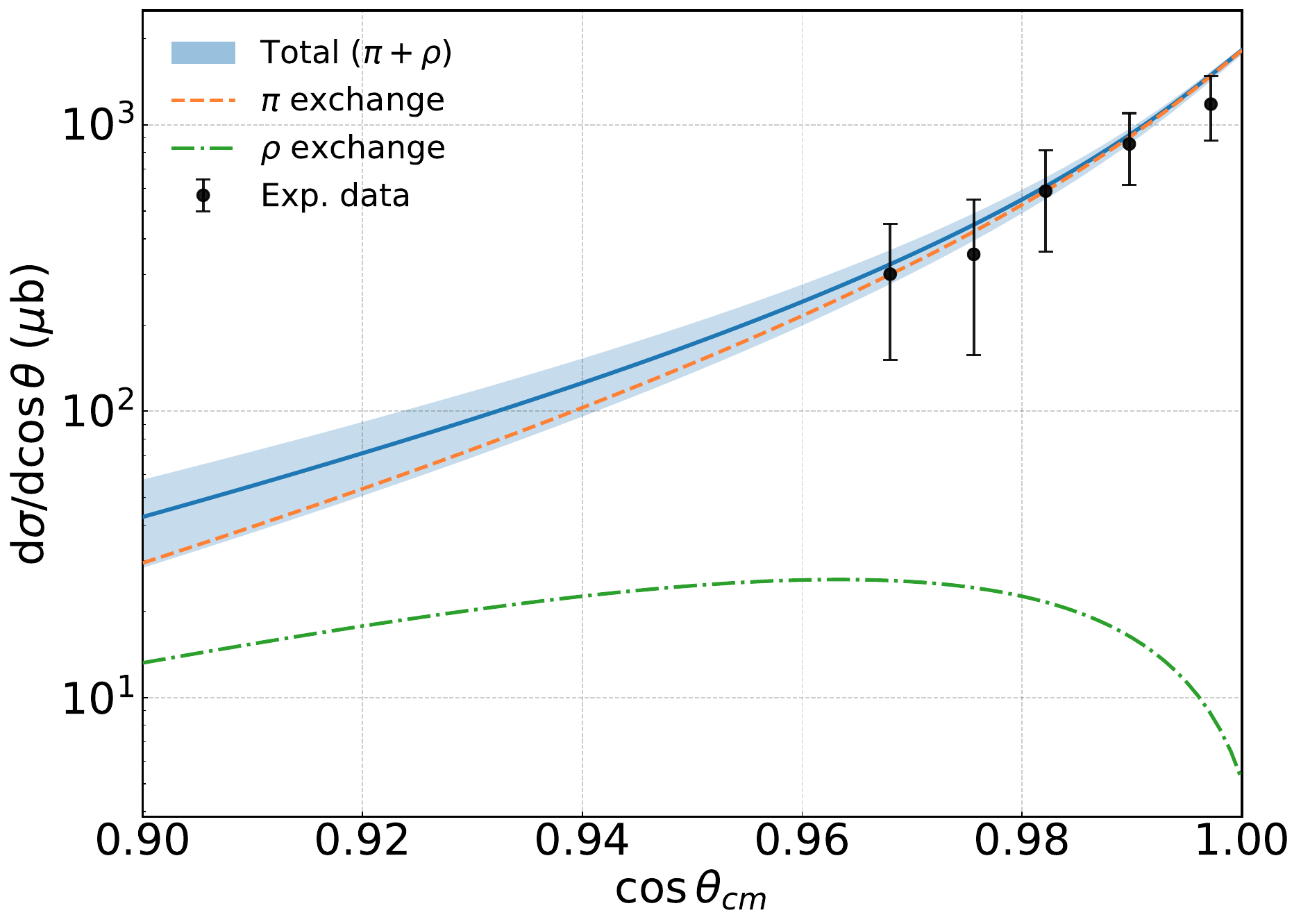}
\caption{The calculated differential production cross section for the $K_3^*(1780)$ state is presented, along with comparisons to experimental data. The differential cross section is evaluated at a center-of-mass energy of 4.08 GeV. The experimental data are taken from Ref.~\cite{Birmingham-CERN-Glasgow-MichiganState-Paris:1984ppi}. The error bands and the solid lines within them represent results obtained with the cutoff parameter set to $1.5\pm 0.2$ GeV. Since the available differential cross section data are concentrated at small forward angles, our theoretical predictions are also presented only in this angular region.}
\label{cs}
\end{figure}

For the second experimental point of the total cross section in Fig.~\ref{ts_total},
corresponding to the center-of-mass energy $\sqrt{s}=4.08~{\rm GeV}$,
differential cross-section data are also available. Based on this
experimental information, we further compare our theoretical result with
the measured angular distribution in Fig.~\ref{cs}. The same
cutoff parameter $\Lambda_t=1.5\pm0.2~{\rm GeV}$, fixed from the total
cross-section fit, is used to evaluate the differential cross section.
The resulting angular distribution agrees well with the experimental data
in the forward-angle region. The calculation also shows a clear
forward-peaked behavior, which is a typical feature of the $t$-channel
exchange process. Such a simultaneous description of both the total and
differential cross sections supports the reliability of the present
effective Lagrangian approach for the $K^{*}_{3}(1780)$ production.

\begin{figure}[htbp]
 \includegraphics[width=0.45\textwidth]{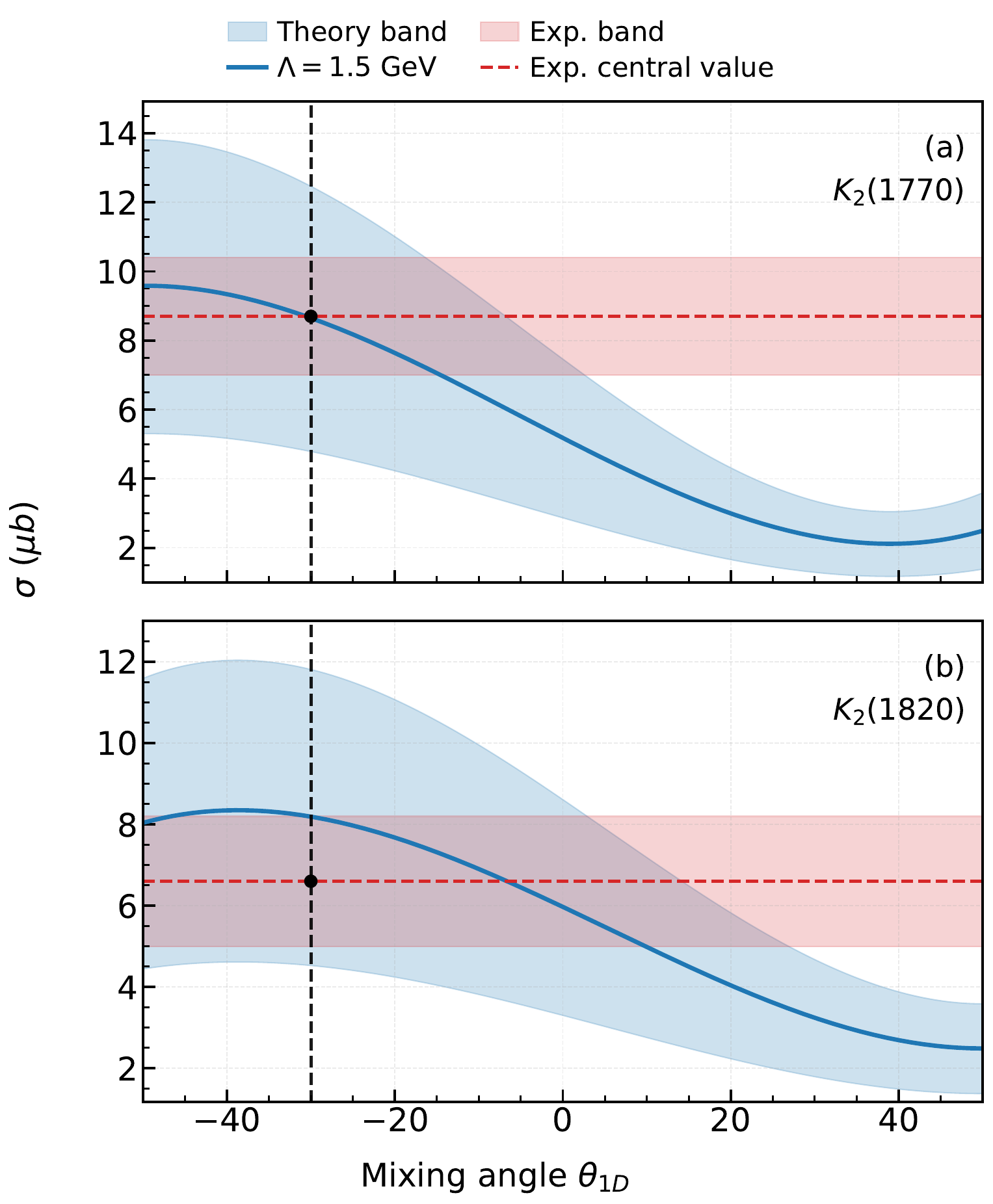}
\caption{
Total cross sections for the reactions $K^- p \to K_2(1770)p$ and
$K^- p \to K_2(1820)p$ as functions of the $K_2(1770)$--$K_2(1820)$
mixing angle. The vertical dashed line denotes the mixing angle
$\theta_{1D}=-30^\circ$ adopted in the numerical calculations. In panel (a),
the red band represents the experimental value
$\sigma[K^-p\to K_2(1770)p]=8.7\pm1.7~\mu{\rm b}$ at
$\sqrt{s}=4.67~{\rm GeV}$. In panel (b), the red band represents the
experimental value
$\sigma[K^-p\to K_2(1820)p]=6.6\pm1.6~\mu{\rm b}$ at the same
center-of-mass energy.
}
\label{mix_1770_1820}
\end{figure}

For the production of the $K_2(1770)$ state, the dependence of the total cross section on
the $K_2(1770)$--$K_2(1820)$ mixing angle is first examined, as shown in Fig.~\ref{mix_1770_1820}.
Since these two physical states arise from the spin mixing between the $1^3D_2$ and $1^1D_2$ configurations, the corresponding three-meson coupling constants are sensitive to the mixing angle through the strong decay widths. Previous studies of their strong decay behaviors suggested that the mixing angle prefers a negative value \cite{Pang:2017dlw}. Motivated by this observation, we calculate the total production cross section as a function of the mixing angle. For the reaction $K^-p\to K_2(1770)p$, the experimental total cross section at $\sqrt{s}=4.67~\mathrm{GeV}$ is $8.7\pm1.7~\mu\mathrm{b}$. The best agreement between the theoretical prediction and the experimental value is obtained around $\theta_{1D}=-30^\circ$, as indicated by the vertical dashed line in Fig.~\ref{mix_1770_1820}(a). This result supports the negative mixing angle favored by the strong decay analysis and provides an independent constraint from the production process.

\begin{figure*}[htbp]
 \includegraphics[width=0.88
    \linewidth]{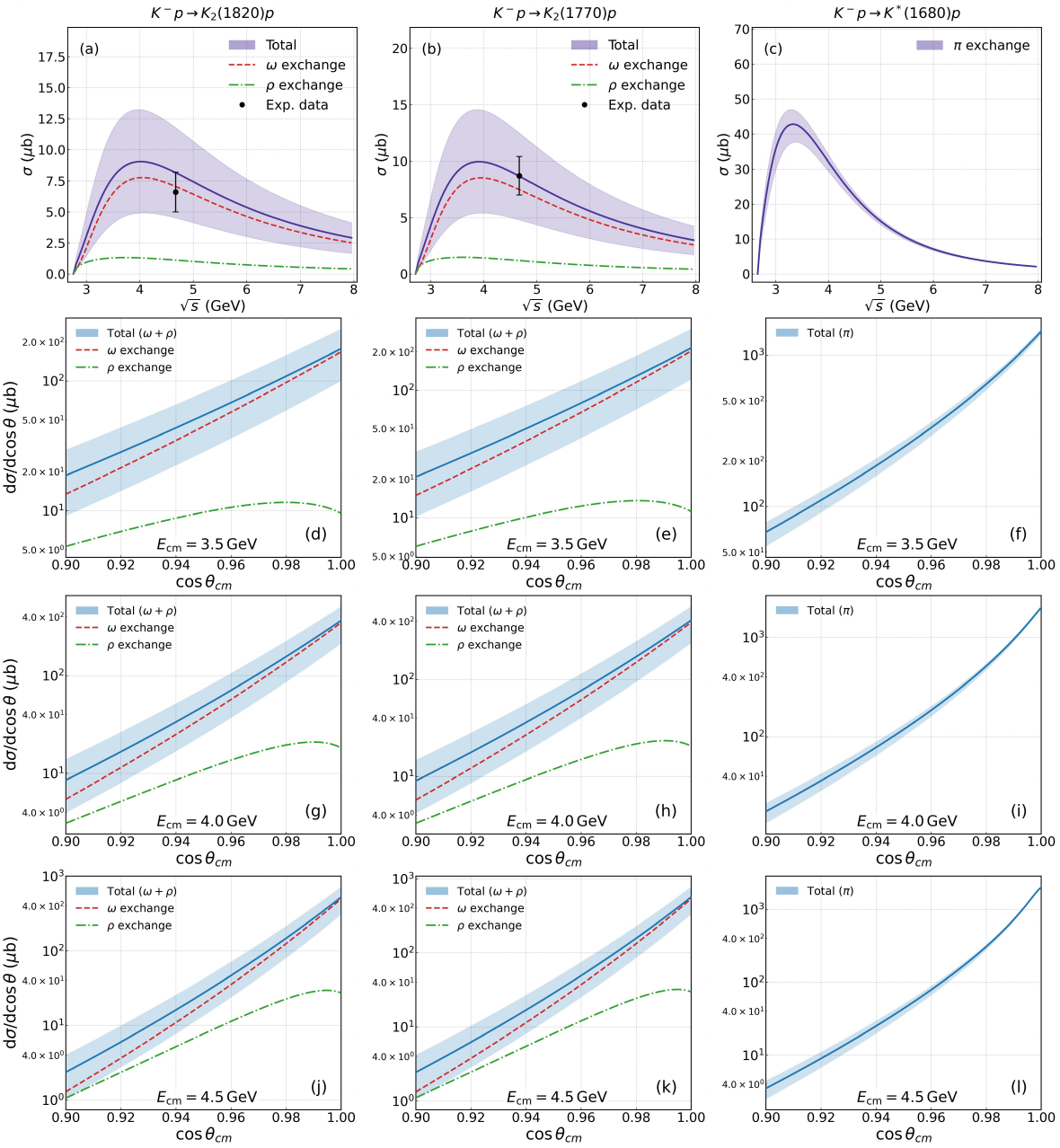}
\caption{The calculated total and differential production cross sections for the $1D$-wave states ($K_2(1820)$, $K_2(1770)$ and $K^*(1680)$) are presented, along with comparisons to experimental data. The upper three panels correspond to the total cross sections, while the lower nine panels show the corresponding differential distributions. The differential cross sections are evaluated at three different center-of-mass energies: 3.5, 4.0, and 4.5 GeV. The experimental data are taken from Refs.~\cite{Aston:1993qc,Estabrooks:1977xe,Birmingham-CERN-Glasgow-MichiganState-Paris:1982tev}. In all figures, the error bands and the solid lines within them represent results obtained with the cutoff parameter set to $1.5\pm 0.2$ GeV. The differential-cross-section panels in the first, second, and third columns correspond to the $K_2(1820)$, $K_2(1770)$ and $K^*(1680)$ mesons, respectively. Because the differential cross sections are concentrated in the small 
forward-angle region, only the angular range close to \(\cos\theta_{cm}=1\) is shown.}
\label{dwavecs}
\end{figure*}

With the same mixing angle $\theta_{1D}=-30^\circ$, we further analyze the total production cross section of the $K_2(1820)$ state. The experimental value for $K^-p\to K_2(1820)p$ at $\sqrt{s}=4.67~\mathrm{GeV}$ is $6.6\pm1.6~\mu\mathrm{b}$. As shown in Fig.~\ref{mix_1770_1820}(b), the theoretical prediction obtained at the same mixing angle also lies within the experimental uncertainty band, indicating that a unified choice of the mixing angle can simultaneously describe the measured total cross sections of both the $K_2(1770)$ and $K_2(1820)$. This agreement is meaningful because the two states are described by the same mixing angle, but their production processes depend on different coupling constants. After fixing the mixing angle, the corresponding coupling constants are used to calculate the energy-dependent total cross sections shown in the left two columns of Fig.~\ref{dwavecs}. In these calculations, both $\rho$- and $\omega$-exchange contributions are included. The $\omega$ exchange gives the dominant contribution, while the $\rho$ exchange provides a smaller correction, and their coherent contribution reproduces the available experimental data reasonably well.

The differential cross sections for the $K^-p\to K_2(1770)p$ reaction are displayed in Fig.~\ref{dwavecs}(e), (h), (k) at several center-of-mass energies. The angular distributions are presented in the forward region, where experimental measurements are available and where the $t$-channel exchange mechanism is expected to be most important. A clear forward-peaked behavior appears as $\cos\theta_{cm}$ approaches unity, which is a typical feature of meson exchange in the $t$ channel. The total differential distribution is mainly governed by the $\omega$-exchange contribution, while the $\rho$-exchange contribution remains smaller over the considered angular range. The uncertainty bands caused by the cutoff parameter variation show that the forward enhancement remains almost unchanged within the model uncertainty. Therefore, the differential behavior of the $K_2(1770)$ production further supports the dominance of the $t$-channel mechanism used in the present calculation.

For the $K^-p\to K_2(1820)p$ reaction, the differential cross sections shown in Fig.~\ref{dwavecs}(d), (g), (j)  exhibit a pattern similar to that of the $K_2(1770)$ state. The distributions increase rapidly toward the forward-angle region, again reflecting the characteristic behavior of the $t$-channel $\rho$- and $\omega$-exchange processes. The $\omega$ exchange provides the leading contribution, whereas the $\rho$ exchange gives a relatively small but visible correction. Since the coupling constants used here are determined at the mixing angle $\theta_{1D}=-30^\circ$, which has already been constrained by the total cross sections of both the $K_2(1770)$ and $K_2(1820)$, the differential predictions in Fig.~\ref{dwavecs} can be regarded as a direct consequence of the same mixing-angle scheme. The simultaneous description of the total and differential cross sections for these two $1D$-wave kaons demonstrates that the present effective Lagrangian approach provides a coherent picture for their production in the $K^-p$ reaction.

For the production of the $K^*(1680)$ state, the calculated total cross section is shown in the rightmost upper panel of Fig.~\ref{dwavecs}(c). In contrast to the $K_2(1770)$ and $K_2(1820)$ cases, only the $\pi$-exchange contribution is retained in the final result for the $K^- p \to K^*(1680)p$ reaction, since the contributions from other possible exchanged mesons are found to be negligibly small.  With the same cutoff parameter $\Lambda_t=1.5\pm0.2~\mathrm{GeV}$ used for the other $1D$-wave states, the total cross section rises rapidly from the threshold region, reaches a sizable maximum at low center-of-mass energies, and then decreases gradually with increasing $\sqrt{s}$. 

The corresponding differential cross sections for the $K^- p \to K^*(1680)p$ reaction are displayed in the rightmost lower panels (f), (i) and (l) of Fig.~\ref{dwavecs} at $\sqrt{s}=3.5$, $4.0$, and $4.5~\mathrm{GeV}$. A pronounced forward enhancement is observed in all three angular distributions, with the differential cross section increasing rapidly as $\cos\theta_{cm}$ approaches unity.  Compared with the $K_2(1770)$ and $K_2(1820)$ cases, the forward-angle differential cross section of the $K^*(1680)$ channel is relatively large, reflecting its strong coupling to the $K\pi$ mode. The uncertainty bands generated by varying the cutoff parameter remain moderate and do not change the forward-peaked structure. Therefore, the combined description of the total and differential cross sections suggests that the present effective Lagrangian framework gives a consistent account of the $K^*(1680)$ production mechanism.

\subsection{Analysis of $1F$-wave kaon production} 

\begin{figure*}[htbp]
 \includegraphics[width=1.
    \linewidth]{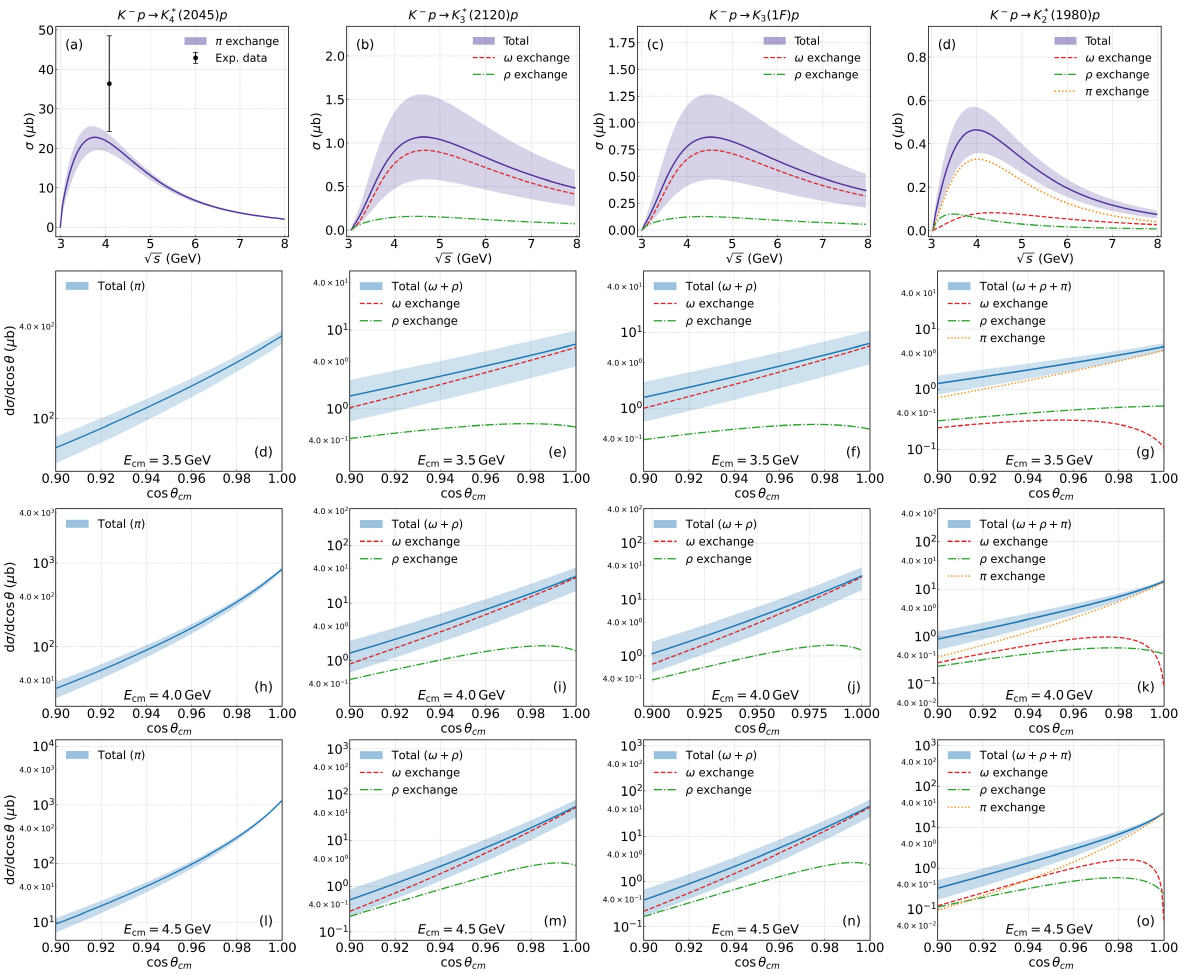}
\caption{The calculated total and differential production cross sections for the $1F$-wave states ($K_4^*(2045)$, $K_3^{\prime}(2120)$, $K_3(1F)$  and $K_2^*(1980)$) are presented. The upper four panels correspond to the total cross sections, while the lower twelve panels show the corresponding differential distributions. The differential cross sections are evaluated at three different center-of-mass energies: 3.5, 4.0, and 4.5 GeV. The experimental data are taken from Ref.~\cite{Birmingham-CERN-Glasgow-MichiganState-Paris:1982tev}. In all figures, the error bands and the solid lines within them represent results obtained with the cutoff parameter set to $1.5\pm 0.2$ GeV. The four differential cross section panels in the first, second, third and fourth columns correspond to the ($K_4^*(2045)$, $K_3^{\prime}(2120)$, $K_3(1F)$  and $K_2^*(1980)$ mesons, respectively. Because the differential cross sections are concentrated in the small 
forward-angle region, only the angular range close to \(\cos\theta_{cm}=1\) is shown.}
\label{fwavecs}
\end{figure*}

For the $K^-p\to K^{\ast}_4(2045)p$ reaction, the production is evaluated through the $t$-channel $\pi$-exchange mechanism. As shown in the first column of Fig.~\ref{fwavecs}, the calculated total cross section rises rapidly from the threshold region and reaches a maximum of about $25.6~\mu\mathrm{b}$, after which it decreases gradually with increasing $\sqrt{s}$. The only available experimental measurement is located at $\sqrt{s}=4.08~\mathrm{GeV}$, with $\sigma_{\mathrm{exp}}=36.4\pm12.1~\mu\mathrm{b}$ \cite{Birmingham-CERN-Glasgow-MichiganState-Paris:1982tev}. Although the theoretical curve is lower than the central value of this experimental point, the predicted value lies close to the lower edge of the experimental uncertainty band.

The corresponding differential cross sections for $K^-p\to K^{\ast}_4(2045)p$ are calculated at $\sqrt{s}=3.5$, $4.0$, and $4.5~\mathrm{GeV}$, as displayed in the lower panels of the first column of Fig.~\ref{fwavecs}. A pronounced forward-peaked behavior appears at all three energies, with $d\sigma/d\cos\theta_{cm}$ increasing rapidly as $\cos\theta_{cm}$ approaches unity.  At $\sqrt{s}=3.5~\mathrm{GeV}$, the distribution already shows a clear forward enhancement and reaches about $10^2~\mu\mathrm{b}$ near the forward end. When the energy increases to $\sqrt{s}=4.0~\mathrm{GeV}$, the angular distribution becomes steeper, and the forward differential cross section is further enhanced. At $\sqrt{s}=4.5~\mathrm{GeV}$, although the total cross section has started to decrease, the differential cross section is more strongly concentrated in the very forward region and reaches the order of $10^3~\mu\mathrm{b}$ as $\cos\theta_{cm}\to1$. Therefore, the production of the $K^{\ast}_4(2045)$ is expected to be most visible in the forward-angle region, which provides a useful guide for future measurements of this $1F$-wave kaon state in $K^-p$ scattering experiments \cite{WANG:2025fmh}.

The recently observed $K_3^{\prime}(2120)$ state deserves particular attention, since the available experimental analysis has established its resonance parameters but has not yet provided information on its production cross section~\cite{COMPASS:2025wkw}. The decay calculation supports its assignment as a $1F$-wave kaon. The predicted total width is consistent with the experimental result~\cite{COMPASS:2025wkw}. The present production analysis therefore provides complementary information that can be used to design a dedicated search in the $K^-p$ reaction. As shown in Fig.~\ref{fwavecs}(b), the total cross section for $K^-p\to K_3^{\prime}(2120)p$ rises rapidly above threshold and develops a broad maximum around $\sqrt{s}\simeq 4.5~\mathrm{GeV}$. Within the adopted cutoff variation, the upper edge of the theoretical band reaches approximately $1.6~\mu\mathrm{b}$, while the central curve is close to $1.1~\mu\mathrm{b}$ at the maximum. The decomposition of the result indicates that the $\omega$-exchange contribution determines the overall magnitude and energy dependence, whereas the $\rho$ exchange provides a smaller correction. The differential distributions in Fig.~\ref{fwavecs}(e), Fig.~\ref{fwavecs}(i), and Fig.~\ref{fwavecs}(m) reveal an additional feature. At $\sqrt{s}=3.5~\mathrm{GeV}$, the cross section already increases toward the forward direction and reaches several $\mu\mathrm{b}$ as $\cos\theta_{cm}\to 1$. At $\sqrt{s}=4.0~\mathrm{GeV}$, the forward yield increases to the order of $10~\mu\mathrm{b}$, and at $\sqrt{s}=4.5~\mathrm{GeV}$ it reaches several tens of $\mu\mathrm{b}$ in the most forward angular region. Meanwhile, the enhancement away from the extreme forward region is much less pronounced.  Consequently, a measurement with sufficient acceptance close to $\cos\theta_{cm}=1$ would provide a useful and experimentally testable strategy for confirming the $K_3^{\prime}(2120)$ state.

The unobserved $K_3(1F)$ state is the spin-mixing partner of the $K_3^{\prime}(2120)$ and can be investigated within the same experimental setup.  The production result in Fig.~\ref{fwavecs}(c) exhibits an energy dependence similar to that of the $K_3^{\prime}(2120)$. The total cross section reaches a broad maximum in the same energy region, with the upper edge of the theoretical uncertainty band approaching $1.25~\mu\mathrm{b}$ and the central prediction remaining around $0.9~\mu\mathrm{b}$. As in the $K_3^{\prime}(2120)$ channel, the $\omega$ exchange gives the leading contribution, whereas the $\rho$-exchange component is subdominant. The differential cross sections displayed in Fig.~\ref{fwavecs}(f), Fig.~\ref{fwavecs}(j), and Fig.~\ref{fwavecs}(n) show that the angular distribution becomes increasingly concentrated near the forward limit as the center-of-mass energy is raised from $3.5$ to $4.5~\mathrm{GeV}$. Near $\cos\theta_{cm}=1$, the predicted differential cross section increases from several $\mu\mathrm{b}$ at $\sqrt{s}=3.5~\mathrm{GeV}$ to the order of $10~\mu\mathrm{b}$ at $\sqrt{s}=4.0~\mathrm{GeV}$, and then to several tens of $\mu\mathrm{b}$ at $\sqrt{s}=4.5~\mathrm{GeV}$. The close similarity between the angular distributions of the two mixed $1F$ states implies that the production angle alone may not be sufficient to separate them. A complementary analysis of the invariant-mass spectra and their  decay properties would provide a more discriminating test of the proposed spin-mixing scenario.

For the $K^{-}p\to K^{*}_{2}(1980)p$ reaction, the predicted total cross section is shown in the upper-right panel of Fig.~\ref{fwavecs}(d). In this calculation, the $\omega$-, $\rho$-, and $\pi$-exchange mechanisms are all taken into account, which makes this channel different from several other $1F$-wave states where only one or two exchanged mesons give sizable contributions. The total cross section rises rapidly from threshold and reaches its maximum value around $\sqrt{s}\simeq 3.8$--$4.0~\mathrm{GeV}$, with the central theoretical peak being about $0.5~\mu\mathrm{b}$. After the peak, the cross section decreases gradually as the center-of-mass energy increases. The decomposition of the total result indicates that the $\pi$-exchange contribution is the largest one, while the $\omega$ and $\rho$ exchanges also give visible but smaller contributions. This feature is especially interesting because, as shown in Table~\ref{width}, the partial width of the $K^{*}_{2}(1980)\to \pi K$ channel is much smaller than those of the $\rho K$ and $\omega K$ channels. Therefore, the production strength is not determined only by the decay width; it is also strongly affected by the light pion propagator, the $\pi NN$ coupling, and the momentum dependence of the $t$-channel amplitude. Compared with the $1D$-wave kaon production shown in Fig.~\ref{dwavecs}, the overall magnitude of the $1F$-wave kaon production is evidently reduced, which is consistent with the general tendency that the production cross section becomes smaller when the orbital angular momentum of the produced state increases.

The corresponding differential cross sections for $K^{-}p\to K^{*}_{2}(1980)p$ are displayed in the rightmost lower panels of Fig.~\ref{fwavecs}(g), Fig.~\ref{fwavecs}(k), and Fig.~\ref{fwavecs}(o) at $\sqrt{s}=3.5$, $4.0$, and $4.5~\mathrm{GeV}$, respectively. At all three energies, the angular distributions exhibit a clear forward enhancement. At $\sqrt{s}=3.5~\mathrm{GeV}$, the differential cross section is relatively moderate and reaches only several $\mu\mathrm{b}$ in the very forward region. When the energy increases to $\sqrt{s}=4.0~\mathrm{GeV}$, the angular distribution becomes steeper, and the forward differential cross section increases to the order of $10~\mu\mathrm{b}$. At $\sqrt{s}=4.5~\mathrm{GeV}$, although the total cross section has already started to decrease from its peak region, the differential distribution is more concentrated near $\cos\theta_{cm}=1$, and the forward value remains at the order of $10~\mu\mathrm{b}$. In these three panels, the $\pi$-exchange contribution grows most rapidly toward the forward direction and dominates the total result, while the $\omega$- and $\rho$-exchange contributions remain non-negligible corrections. Thus, the forward-angle region provides the most favorable kinematic window for searching for the $K^{*}_{2}(1980)$ production in future $K^{-}p$ scattering measurements \cite{WANG:2025fmh}.

\subsection{Analysis of $1G$-wave kaon production} 

\begin{figure*}[htbp]
 \includegraphics[width=1.
    \linewidth]{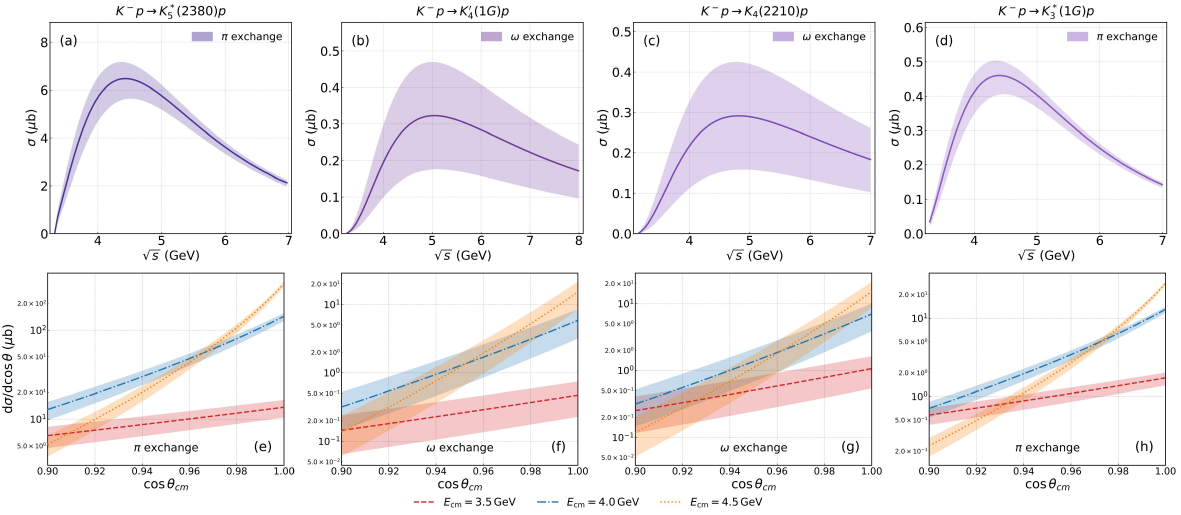}
\caption{
Calculated total and differential production cross sections for the 
\(1G\)-wave strange meson states 
\(K_5^*(2380)\), \(K_4'(1G)\), \(K_4(2210)\), and \(K_3^*(1G)\) 
in the \(K^-p\) reaction. 
Panels (a)--(d) show the total cross sections as functions of the 
center-of-mass energy \(\sqrt{s}\), while panels (e)--(h) present the 
corresponding differential cross sections \(d\sigma/d\cos\theta_{cm}\) 
in the forward-angle region. 
The differential cross sections are evaluated at 
\(E_{\rm cm}=3.5\), \(4.0\), and \(4.5~{\rm GeV}\). 
From left to right, the four columns correspond to the production of 
\(K_5^*(2380)\), \(K_4'(1G)\), \(K_4(2210)\), and \(K_3^*(1G)\), respectively. 
The shaded bands denote the theoretical uncertainties induced by varying 
the cutoff parameter within \(\Lambda=1.5\pm0.2~{\rm GeV}\), while the solid 
curves represent the central predictions with \(\Lambda_t=1.5~{\rm GeV}\). 
Because the differential cross sections are concentrated in the small 
forward-angle region, only the angular range close to \(\cos\theta_{cm}=1\) is shown.
}
\label{gwavecs}
\end{figure*}

For the $K^{-}p\to K^{*}_{5}(2380)p$ reaction, the production process is dominated by the $t$-channel $\pi$-exchange mechanism, as shown in Fig.~\ref{gwavecs}(a). The calculated total cross section increases rapidly from the threshold region and reaches its maximum value of about $6.5~\mu{\rm b}$ around $\sqrt{s}\simeq 4.4~{\rm GeV}$. After the peak, the cross section decreases gradually with increasing center-of-mass energy and decreases to about $2~\mu{\rm b}$ at $\sqrt{s}=7.0~{\rm GeV}$. Among the four $1G$-wave states considered here, the $K^{*}_{5}(2380)$ channel gives the largest total cross section. Nevertheless, compared with the $1D$- and $1F$-wave kaon productions discussed above, its magnitude is still reduced. This behavior supports the general tendency that the production cross section decreases as the orbital angular momentum of the produced kaon state increases, namely the overall hierarchy $\sigma_{1G}<\sigma_{1F}<\sigma_{1D}$.

The corresponding differential cross section for $K^{-}p\to K^{*}_{5}(2380)p$ is displayed in Fig.~\ref{gwavecs}(e), where the results at $\sqrt{s}=3.5$, $4.0$, and $4.5~{\rm GeV}$ are shown together. A clear forward-peaked structure appears at all three energies, which is the typical feature of the $t$-channel exchange process. Near $\cos\theta_{cm}=1$, the differential cross section is only at the order of $10~\mu{\rm b}$ for $\sqrt{s}=3.5~{\rm GeV}$, while it increases to the order of $10^{2}~\mu{\rm b}$ at $\sqrt{s}=4.0~{\rm GeV}$. At $\sqrt{s}=4.5~{\rm GeV}$, where the total cross section is close to its peak region, the angular distribution becomes much steeper and the forward differential cross section can reach several $10^{2}~\mu{\rm b}$. Therefore, although the integrated production rate of this $1G$-wave state is smaller than those of the lower-orbital kaons, the very forward region still provides a favorable kinematic window for searching for the $K^{*}_{5}(2380)$ in future $K^{-}p$ measurements~\cite{WANG:2025fmh}.

For the $K^{-}p\to K'_{4}(1G)p$ reaction, the calculated total cross section is presented in Fig.~\ref{gwavecs}(b). In this channel, only the $\omega$-exchange contribution is retained in the final calculation, because the other possible exchanged-meson contributions are found to be strongly suppressed. This is different from the mixed $1D$-wave states $K_{2}(1770)$ and $K_{2}(1820)$, as well as the mixed $1F$-wave states $K'_{3}(2120)$ and $K_{3}(1F)$, where both $\rho$- and $\omega$-exchange mechanisms were included. The total cross section of $K'_{4}(1G)$ rises from threshold and develops a broad maximum around $\sqrt{s}\simeq 5.0~{\rm GeV}$. The central value of the peak is approximately $0.32~\mu{\rm b}$, with a sizable uncertainty band induced by the cutoff parameter. This magnitude is much smaller than that of the $K^{*}_{5}(2380)$ channel and is also below the typical production scale of the $1F$-wave mixed states, reflecting the suppression associated with the higher orbital excitation.

The differential cross section for $K^{-}p\to K'_{4}(1G)p$ is shown in Fig.~\ref{gwavecs}(f). The angular distributions at $\sqrt{s}=3.5$, $4.0$, and $4.5~{\rm GeV}$ all increase toward the forward direction. At $\sqrt{s}=3.5~{\rm GeV}$, the forward differential cross section is relatively small and is below $1~\mu{\rm b}$ near $\cos\theta_{cm}=1$. When the energy is raised to $\sqrt{s}=4.0~{\rm GeV}$, the forward value increases to several $\mu{\rm b}$. At $\sqrt{s}=4.5~{\rm GeV}$, the distribution becomes more strongly concentrated near $\cos\theta_{cm}=1$, and the forward differential cross section reaches the order of $10~\mu{\rm b}$. 

For the newly observed $K_{4}(2210)$ state, the total production cross section of the $K^{-}p\to K_{4}(2210)p$ reaction is shown in Fig.~\ref{gwavecs}(c). Since the $K_{4}(2210)$ is a newly reported strange meson, its production prediction is particularly important for future confirmation of this state in kaon-beam experiments~\cite{WANG:2025fmh}. Similar to its mixed partner $K'_{4}(1G)$, only the $\omega$-exchange mechanism is included in the calculation, because the $\rho$-exchange and other possible contributions are numerically very small. The predicted total cross section of the $K_{4}(2210)$ reaches a broad maximum of about $0.29~\mu{\rm b}$ around $\sqrt{s}\simeq 4.8$--$5.0~{\rm GeV}$ and then decreases slowly as $\sqrt{s}$ increases. Its production strength is comparable to that of $K'_{4}(1G)$.

The corresponding angular distributions for $K^{-}p\to K_{4}(2210)p$ are displayed in Fig.~\ref{gwavecs}(g). At $\sqrt{s}=3.5~{\rm GeV}$, the differential cross section is relatively weak over the whole forward-angle region and reaches about $1~\mu{\rm b}$ near $\cos\theta_{cm}=1$. At $\sqrt{s}=4.0~{\rm GeV}$, the forward yield increases to several $\mu{\rm b}$. When the energy becomes $\sqrt{s}=4.5~{\rm GeV}$, the angular distribution grows more rapidly as $\cos\theta_{cm}$ approaches unity, and the forward value reaches the order of $10~\mu{\rm b}$. The curves at different energies also show that the higher-energy distribution is more strongly compressed into the very forward region. 

For the $K^{-}p\to K^{*}_{3}(1G)p$ reaction, the total cross section is plotted in Fig.~\ref{gwavecs}(d). The production is evaluated through the dominant $\pi$-exchange mechanism, while the contributions from other possible exchanged mesons are neglected because they are much smaller in the numerical calculation. The total cross section rises quickly from threshold and reaches a maximum of about $0.46~\mu{\rm b}$ around $\sqrt{s}\simeq 4.4~{\rm GeV}$. After the peak, the cross section decreases continuously with increasing energy. 

The differential cross section for $K^{-}p\to K^{*}_{3}(1G)p$ is presented in Fig.~\ref{gwavecs}(h). Similar to the other $1G$-wave channels, the angular distributions at $\sqrt{s}=3.5$, $4.0$, and $4.5~{\rm GeV}$ exhibit an evident forward enhancement. At $\sqrt{s}=3.5~{\rm GeV}$, the forward differential cross section is only at the level of a few $\mu{\rm b}$. At $\sqrt{s}=4.0~{\rm GeV}$, it increases to the order of $10~\mu{\rm b}$ near $\cos\theta_{cm}=1$. At $\sqrt{s}=4.5~{\rm GeV}$, the distribution becomes the steepest among the three energies and the forward value can reach several tens of $\mu{\rm b}$. 

\section{Summary}
\label{section4}

In this work, we have systematically investigated the production of high-orbital kaon excitations in the $K^- p$ reaction, with emphasis on the $1D$-, $1F$-, and $1G$-wave strange-meson states. Within the effective Lagrangian approach, the dominant $t$-channel exchange mechanisms were constructed for the relevant production processes. The exchanged mesons were selected according to the dominant kaon-containing decay channels of the produced states. In addition, the finite-size effects at the interaction vertices were taken into account by introducing a phenomenological form factor. The cutoff parameter was determined to be $\Lambda_t = 1.5 \pm 0.2~\mathrm{GeV}$ by fitting the available experimental data for the $K^- p \to K_3^*(1780)p$ reaction. With the same value of $\Lambda_t$, the production of other high-orbital kaon states was further studied without introducing additional free parameters.

For the $1D$-wave kaons, the available experimental data for
the $K_3^*(1780)$, $K_2(1770)$, and $K_2(1820)$ are used to constrain and
test the present model. The production of the $K^*(1680)$ is then evaluated
with the same parameter set, and its total cross section and angular
distributions are presented as theoretical predictions, since no direct
experimental production cross-section data are available in the
considered energy region.

Using the same theoretical framework, we further predicted the production cross sections of the $1F$- and $1G$-wave kaon states. The obtained result for the $K_4^*(2045)$ is compatible with the available experimental information. Predictions were also made for the $K_3^\prime(2120)$, $K_3(1F)$, $K_2^*(1980)$, $K_5^*(2380)$, $K_4^\prime(1G)$, $K_4(2210)$, and $K_3^*(1G)$. These results provide useful theoretical guidance for the search for poorly established or still-unobserved high-orbital kaon states. Although the production strength generally decreases with increasing orbital excitation, the predicted cross sections remain sizable. Moreover, the differential cross sections are mainly concentrated in the forward-angle region, suggesting that forward-angle measurements are especially favorable for observing these excited strange mesons.

Overall, the present study provides a unified description of high-orbital kaon production in $K^- p$ scattering within the effective Lagrangian approach. The successful reproduction of existing experimental data and the extension to the less-explored $1F$- and $1G$-wave sectors demonstrate the reliability and predictive power of this framework. These results not only improve our understanding of the production dynamics of excited kaons, but also provide valuable theoretical references for future kaon-beam experiments at facilities such as J-PARC and other high-intensity meson-beam programs.

\vfil

\begin{acknowledgments}

 This work is supported by the Natural Science Foundation of Gansu Province (No. 26RCKA012, No. 25JRRA799), the National Natural Science Foundation of China under Grants No. 12335001 and No. 12247101, the ‘111 Center’ under Grant No. B20063, the Fundamental Research Funds for the Central Universities (No. lzujbky-2023-stlt01), and Lanzhou City High-Level Talent Funding.

\end{acknowledgments}

\vfil

\bibliographystyle{apsrev4-1}
\bibliography{ref}

\end{document}